\documentclass[aps,prd,onecolumn,showpacs,groupedaddress]{revtex4-2}
\usepackage{graphicx}
\usepackage{dcolumn}
\usepackage{bm}
\usepackage{color}
\usepackage{supertabular}
\usepackage[T1]{fontenc}
\usepackage{epstopdf}
\usepackage{amsmath}
\usepackage{amssymb}
\usepackage{setspace}
\usepackage{multirow}
\usepackage{booktabs}
\usepackage[section]{placeins}
\usepackage{mathrsfs}
\usepackage{hyperref}
\usepackage{caption}
\usepackage{subcaption}
\usepackage{float}

\makeatletter

\newcommand{\Rmnum}[1]{\expandafter\@slowromancap\romannumeral #1@}

\makeatother

\begin{document}

\title{On sensitivity of nucleus deformation on final-state flow harmonics}

\author{Henrique Mascalhusk$^{1,2}$}
\author{Dener S. Lemos$^{3,4}$}
\author{Otavio Socolowski Jr.$^{5}$}
\author{Wei-Liang Qian$^{6,2,1}$}\email[E-mail: ]{wlqian@usp.br}
\author{Sandra S. Padula$^{4}$}
\author{Rui-Hong Yue$^{1}$}

\affiliation{$^{1}$ Center for Gravitation and Cosmology, College of Physical Science and Technology, Yangzhou University, Yangzhou 225009, China}
\affiliation{$^{2}$ Faculdade de Engenharia de Guaratinguet\'a, Universidade Estadual Paulista, 12516-410, Guaratinguet\'a, SP, Brazil}
\affiliation{$^{3}$ Department of Physics, University of Illinois at Chicago, Chicago, Illinois 60607, USA}
\affiliation{$^{4}$ Instituto de F\'isica Te\'orica, Universidade Estadual Paulista, 01140-070, S\~ao Paulo, SP, Brazil}
\affiliation{$^{5}$ Instituto de Matem\'atica Estat\'istica e F\'isica, Universidade Federal do Rio Grande, 96201-900 Rio Grande, RS, Brazil}
\affiliation{$^{6}$ Escola de Engenharia de Lorena, Universidade de S\~ao Paulo, 12602-810, Lorena, SP, Brazil}

\begin{abstract}
In this work, we explore the effect of deformation of the nuclei on collective flow in relativistic heavy-ion collisions.
The parameter associated with the geometrical deformation in the Glauber model is tuned to reproduce the empirical multiplicity probability distributions correctly.
Subsequently, the particle spectra and collective flows for Au+Au and U+U collisions are evaluated using a hybrid hydrodynamic code CHESS.
We analyze the effects of the degrees of freedom associated with the IC on the final-state flow harmonics by exploring the parameter space of the former.
The connection between the deformation parameters, specifically $\beta_2$ and $\beta_4$, and the flow anisotropies is scrutinized.
In particular, deviations in elliptic flow at $p_{\mathrm{T}}\sim 2$ GeV are observed at smaller values of $\beta_2$ in Au+Au collisions.
On the other hand, for U+U collisions, the averaged overall flow harmonics are found to be less sensitive to the geometrical parameters.
Despite the difference in the model's specifications, our findings largely confirm those obtained in the literature employing different approaches, which indicate that flow harmonics can be used as a sensible probe for the initial geometry fluctuations and to discriminate between different theoretical models.
\end{abstract}

\date{Dec 14th, 2024}

\maketitle

\newpage

\section{Introduction}

The relativistic heavy-ion collision is one of the crucial tools to extract crucial information on the quark-gluon plasma (QGP)~\cite{Collins, Cabibbo, Rafelski}.
As a state of matter subject to extreme conditions, QGP is featured by high temperature and density created at the moment of collisions.
On the experimental side, among others, the collective phenomenon, notably the elliptic flow, plays a significant role in forming the community's consensus on the fluid nature of the strongly interacting system. 
The latter is demonstrated in terms of the observations of nuclear collisions carried out at the relativistic heavy-ion collider~\cite{Ackermann} (RHIC) and large hadron collider~\cite{Aamodt, Shen, ALICE} (LHC), inclusively for ``small'' systems~\cite{Werner, NoronhaA, daSilva2022xwu, Hippert2020sat, Hippert:2020kde}.

While applicable to physical systems of distinct scales, the transport, particularly the hydrodynamic model, furnishes an effective and intuitive description for the temporal evolution of the system produced at heavy-ion collision experiments~\cite{hydro-review-04,sph-review-01,hydro-review-06,hydro-review-08,hydro-review-10,NoronhaB}.
In practice, it has become one of the standard tools for analyzing the relevant experimental data~\cite{Adare2016, sph-cfo-01, sph-corr-13, Das, sph-review-02}.
A complete and systematic approach to the collision process mostly consists of four stages: pre-collision, collision, temporal evolution, and decoupling.
Numerical implementations of the underlying theoretical models have been extensively developed by the community.
The latter typically involves the initial conditions (IC)~\cite{Trento, Schenke}, the relativistic hydrodynamical evolution~\cite{vHLLE, MUSIC}, the hadronization and final-state interactions~\cite{THERMINATOR}. 
Using numerical simulations, the results can be compared against the measurements from various experimental collaborations at RHIC and LHC.

Recently, the implications of the nucleus' geometric deformation have aroused much interest. 
It was speculated~\cite{Heyde2011, Heyde2016} that such a deformation might have a sizable effect on some experimental observables. 
To be specific, described by a few parameters, the deviation from isotropic nuclei is shown to give rise to a relevant impact on the initial eccentricity~\cite{Filip}, momentum-dependent particle correlations, and collective flow~\cite{Samanta}, particularly the elliptic flow. 

Using the Glauber Monte Carlo approach, the eccentricities of the IC for Au+Au and U+U collisions were first studied in~\cite{Filip}. 
The relation between the initial state geometry and fluctuations was investigated~\cite{Schenke1} for Au+Au, Cu+Au, and U+U collisions by employing the Glauber and IP-Glasma approaches.
It was concluded that elliptic flow as a function of multiplicity in central collisions can be used to discriminate between the models with qualitatively different particle production mechanisms.
On the experimental side, efforts were made~\cite{star} by STAR Collaboration to explore such an effect.
By selecting entirely overlapping collisions of Au+Au and at $\sqrt{s_{NN}}=200$ GeV and those of U+U collisions at $\sqrt{s_{NN}}=193$ GeV, the dependence of elliptic flow on multiplicity was measured.
The data seem to favor models for which the multiplicity does not strongly depend on the number of binary nucleon-nucleon collisions.
To estimate the initial system's eccentricities, the density of the initial nucleus shape was modeled by a deformed Woods-Saxon profile.
In particular, when integrating over different nuclear orientations, the Woods-Saxon parameters cease to be appropriate, and the deformation parameter $\beta_2$ derived from the reduced electric quadrupole transition probability does not coincide with that from the spherical harmonic expansion.
The appropriate Woods-Saxon and deformation parameters were evaluated in~\cite{Shou} in a consistent manner.

Also, analyses indicated~\cite{Jia2022Jan, Jia-2022apr} the importance of the multipole deformations. 
Besides $\beta_2$, which measures the quadrupole, such deformation is also characterized by $\beta_3$ and $\beta_4$, namely, the octupole and hexadecapole parameters.
These parameters are relevant for the multipole expansion of the Woods-Saxon distribution for non-spherical nuclei. 
In literature, the values of these parameters typically vary significantly in different studies~\cite{Giacalone, GiacaloneA, GiacaloneB, GiacaloneC, GiacaloneD}.

Moreover, in the past few years, isobar collisions such as Zr+Zr and Ru+Ru were introduced to explore atomic nuclei's skin and shape. 
The quantitative analysis becomes feasible as the relevant bulk observables such as $v_2$ and $v_3$ ratios have been obtained with unprecedented precision~\cite{JiaA, LiA, LiB, Giacalone-Jia, JIA2022137312, Jia-Giacalone-Zhang2023b, Zhang-Bhatta-Jia}. 
Besides, these isobars collisions are distinct in terms of the shapes of the participant nuclei, which are characterized by different deformations.
For the latter, the quadrupole and hexadecapole deformations play significant roles~\cite{Zhang-Jia, Rong}.

In this line of thought, the present study is motivated to investigate the effect of multipole nuclear deformation in flow harmonics.
To this end, we will focus on Au+Au collisions at $\sqrt{s_{NN}}=200$ GeV and U+U collisions at $\sqrt{s_{NN}}=193$ GeV.
The non-spherical Woods-Saxon profile will be employed to describe the nuclear density.
We will primarily focus on the quadrupole and hexadecapole deformations in terms of the parameters $\beta_2$ and $\beta_4$.
The hydrodynamic simulations will be carried out using Complete Hydrodynamical Evolution System (CHESS)~\cite{chess_ref}, a hybrid code unifying different stages of the relativistic heavy-ion collisions.
Specifically, the CHESS code is comprised of $T_RENT_o$~\cite{Trento} furnishing the IC, vHLLE~\cite{vHLLE} simulating the hydrodynamic evolution, and THERMINATOR~2~\cite{THERMINATOR} governing the hadronization process. 

The remainder of the paper is organized as follows.
In the following section, we briefly introduce the CHESS code and discuss the specific configuration for the present study.
In Sec~\ref{section3}, we present the numerical results and elaborate, among others, the effect of the initial-state geometric shape on the flow harmonics in terms of the deformation parameters.
The last section is dedicated to further discussions and concluding remarks.

\section{The theoretical framework and the CHESS code}\label{section2}

As mentioned above, there are four main stages for relativistic heavy-ion collisions: pre-collisions, collisions, relativistic hydrodynamics, and decoupling. 
In this section, we elaborate on implementing the above elements in the present study while giving a brief overview of hydrodynamics and the hadronization process.

\subsection{Glauber Model}

The Glauber model~\cite{Glauber} is one of several conventional candidates for the IC.
Following its theoretical framework, nuclear collisions are primarily governed by the underlying geometry, typically given by the Woods-Saxon profile.
Moreover, the Glauber model assumes that the collision between two nuclei can be viewed as a superposition of a series of binary nucleon collisions.
As individual nucleon can be viewed as a pancake, collision only occurs if the distance between the center of two nucleons reach the minimum distance given by
\begin{equation}
\sqrt{\frac{\sigma_{NN}}{\pi}} ,
\end{equation}
where $\sigma_{NN}$ is the total cross-section of elementary nucleon-nucleon collisions.

The two nuclei, denoted by $A$ and $B$, are referred to as the target and projectile, respectively.
They travel in opposite directions along the $z$-axis.
To study the effect of deformed nuclei, besides symmetric ones, we will also consider nuclei subject to quadrapole and hexadecapole deformations, which are parameterized in terms of the deformation parameters in the Woods-Saxon profiles.

Specifically, the Woods-Saxon profile for a symmetric nucleus satisfies
\begin{equation}
    \label{woods-saxon}
    \rho_{A, B}^\mathrm{part} (x, y, z) = \frac{\rho_0}{1 + \exp{\left[\frac{r-R_{A, B}}{a}\right]}},
\end{equation}
where the subscripts $A$ and $B$ denote the two nuclei, $\rho$ is the initial density, $r=\sqrt{x^2 + y^2 + z^2}$, $a$ is the diffusion parameter, and $R_{A, B}$ are the radii of the nuclei.

On the other hand, for a deformed nucleus, we follow the Woods-Saxon profile provided by~\cite{Loizides2014, Giacalone}, given by
\begin{equation}
    \label{woods-saxon-deformed}
    \rho_{A, B}^\mathrm{part} (x, y, z) = \frac{\rho_0}{1 + \exp{\left[\frac{r-R_{A, B}\left(1+\beta_2 Y_{2 0} + \beta_4 Y_{4 0}\right)}{a}\right]}},
\end{equation}
where we have considered quadrupole and hexadecapole deformations governed by the parameters $\beta_2$ and $\beta_4$, and $Y_{2 0}$ and $Y_{4 0}$ are the spherical harmonics.

Recent studies~\cite{Jia2022, Giacalone-Jia} propose different implementations for deformation parameters and their respective associated spherical harmonics according to the degree of deformation used in Eq.~\eqref{woods-saxon-deformed}.
In this work, our primary focus is the impact on the observables originating from the deformations related to parameters $\beta_2$ and $\beta_4$.
By imposing different conditions, we aim to verify the sensitivity of the relevant observables on these deformation parameters.

\begin{figure}[H]
\centering
\includegraphics[width=0.325\linewidth,height=0.25\linewidth]{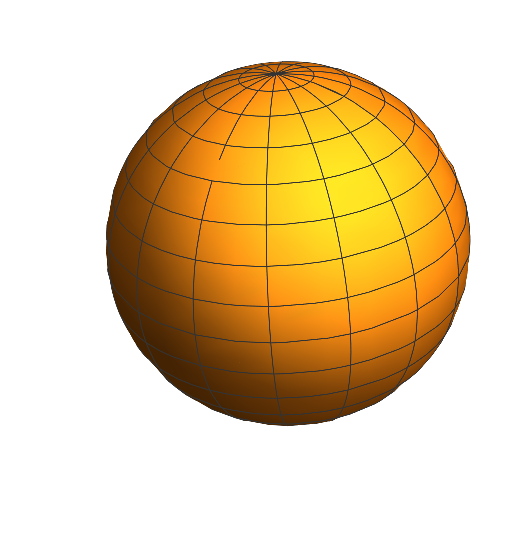} 
\includegraphics[width=0.325\linewidth,height=0.2\linewidth]{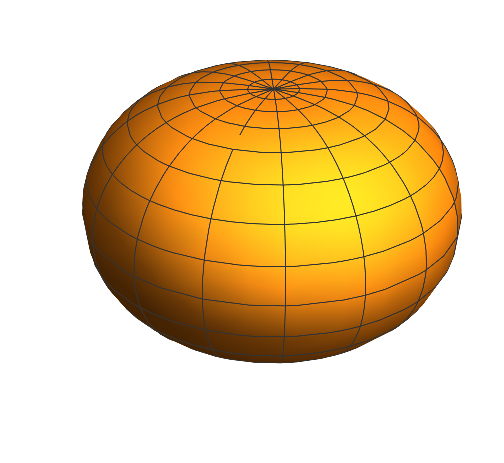} 
\newline
\centering
\includegraphics[width=0.325\linewidth,height=0.25\linewidth]{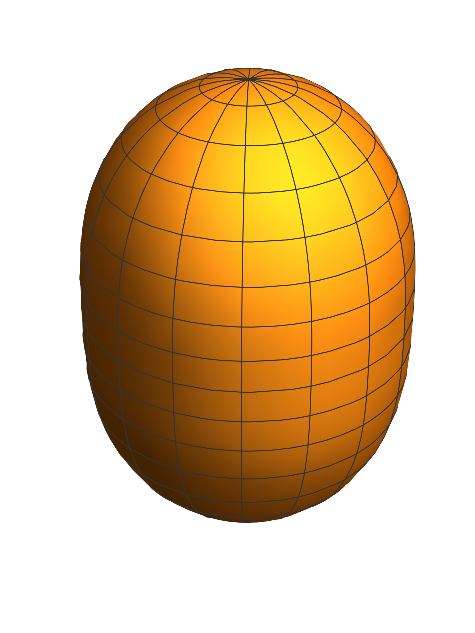} 
\includegraphics[width=0.325\linewidth,height=0.2\linewidth]{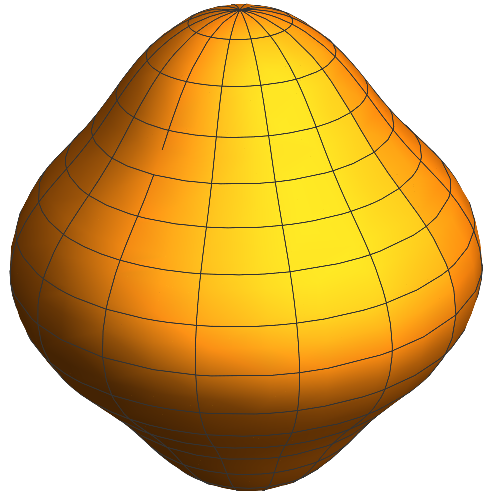}
\newline
\centering
\includegraphics[width=0.48\linewidth,height=0.3\linewidth]{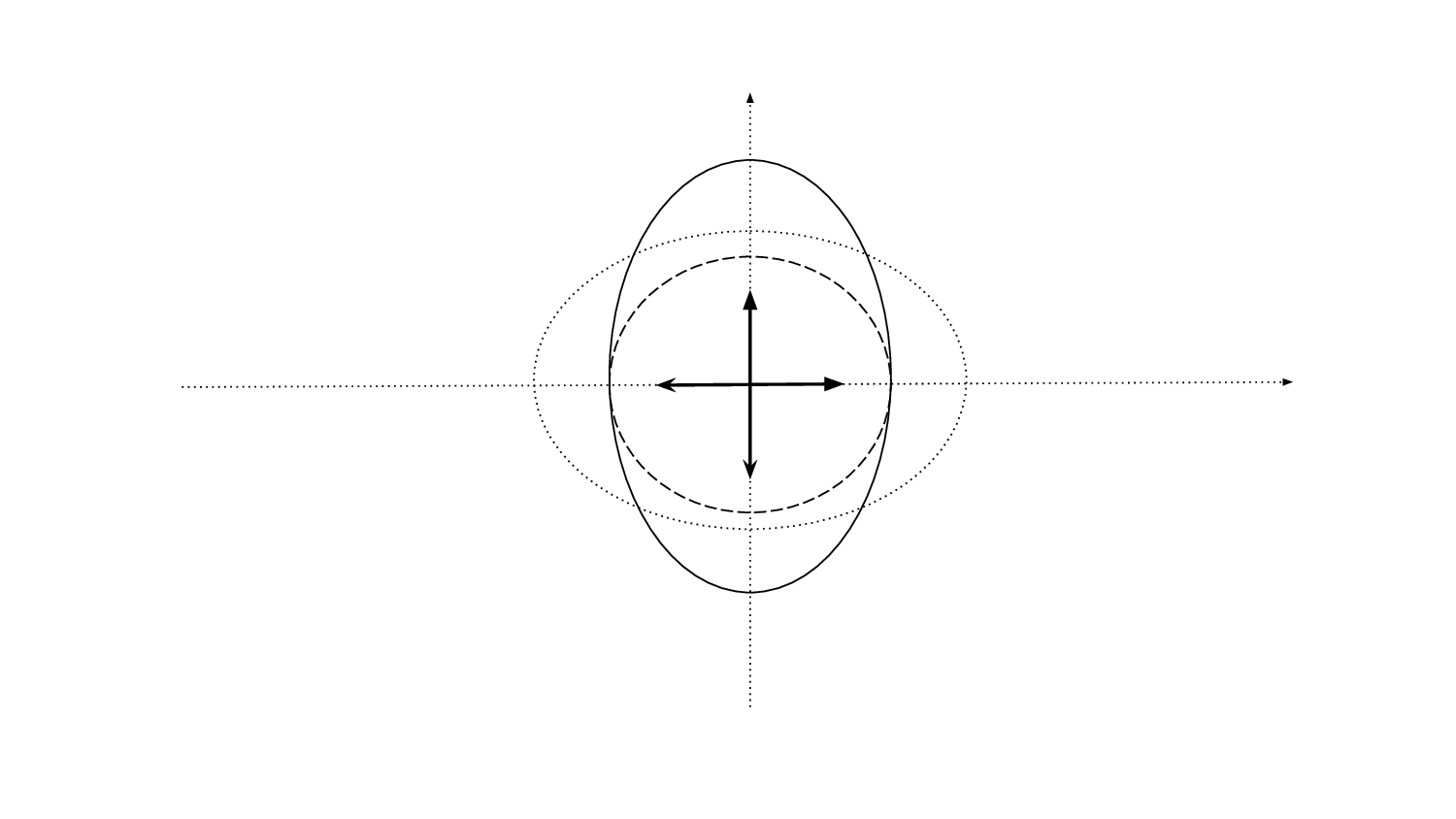} 
\includegraphics[width=0.48\linewidth,height=0.3\linewidth]{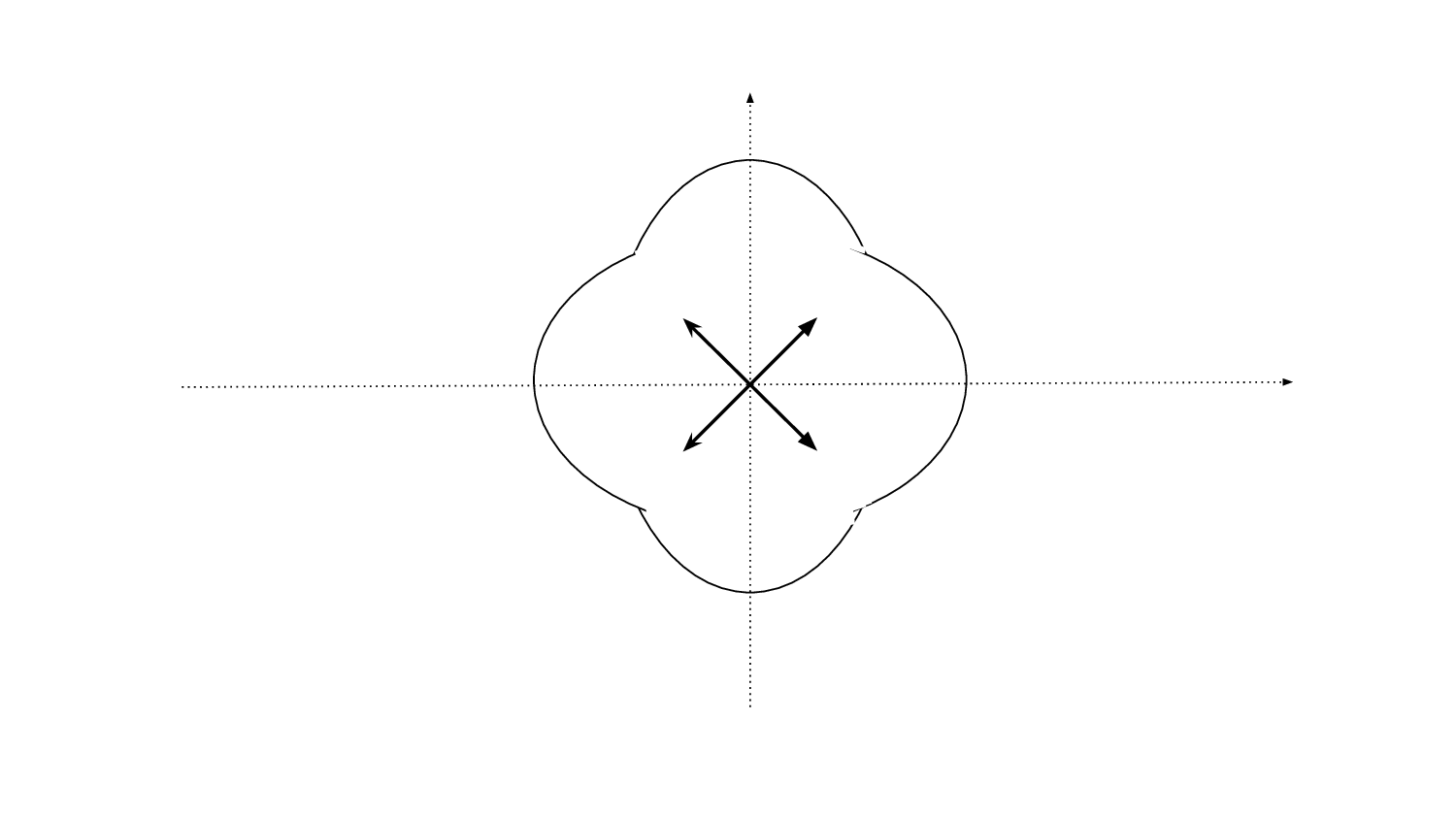}
\caption{Different deformations of the nuclei used as the IC in this study: Top-left: Symmetric sphere representing $\beta_2=\beta_4=0$. 
Top-right: Oblate nucleus owing to a negative $\beta_2$.
Center-left: Prolate nucleus associated with a positive $\beta_2$.
Center-right: Deformed nuclei owing to nonvanishing $\beta_4$. 
Bottom-left: 2D view of deformed nuclei with vanishing $\beta_4$, aiming oblate (dotted line) and prolate (fullfilled line) deformations.
Bottom-right: 2D view of deformed nuclei owing to nonvanishing $\beta_4$.
The arrows indicate the deformation parameters increase.
}
\label{shape}
\end{figure}

Fig.~\ref{shape} illustrates the shapes of the nuclei according to different quadrupole and hexadecapole deformation parameters $\beta_2$ and $\beta_4$. 
The collisions are assumed to occur on the z-axis.
Visually, a variation in $\beta_2$ gives rise to a deformation of essentially oblate or prolate shapes, while the parameter $\beta_4$ implies further ondulation along the zenithal direction, characterizing a shape illustrated in the bottom-right panel of Fig.~\ref{shape}.
The nucleus is spherically symmetric if no deformation parameter is associated with the collision, i.e., $\beta_2=\beta_4=0$.
In the present study, We do not consider deformation degrees of freedom related to rotations in the axis such as those presented in~\cite{Nielsen:2023}.

Using the Wood-Saxon profile of the two nuclei $A$ and $B$, we can also obtain information on nuclear thickness function, differential and total cross sections, the number of nucleon-nucleon collisions, and the total number of participants $N_\mathrm{part}$. 

As a venerable approach, the Glauber model facilitates the study of deformed nuclei owing to its flexibility in encoding the information on deformation via the model parameters.
In the literature, studies have also been carried out using different ICs such as IP-glasma~\cite{Schenke}, an approach based on Color Glass Condensate (CGC)~\cite{McLerran}.

\subsection{Relativistic Hydrodynamics}

Relativistic hydrodynamics have become a prominent tool to describe the collective phenomenon observed in the heavy-ion collisions~\cite{hydro-review-04, sph-review-01, hydro-review-06, hydro-review-07, hydro-review-08, hydro-review-10, sph-review-02}. 
Although the system behaves close to an ideal fluid, viscous hydrodynamics is mainly employed. 
Some pedagogic literature on the specific numerical algorithm can be found in Refs.~\cite{Gourgoulhon, Rezzolla}.

In the present paper, we employ a specific implementation, dubbed vHLLE~\cite{vHLLE}. 
The code uses a subset of the High-Resolution Shock-Capturing (HRSC) algorithms, known as Godunov-type~\cite{Cong}. 
The HRSC algorithm treats the shock waves in their conservation form; it also generalizes to non-conservative shock waves including the treatment of non-continuous changes of variables such as pressure, density, and velocity~\cite{Leer, Shu}. 

The algorithm's mathematical framework consists of using solutions to Riemann problems that consider cell boundaries. 
The constituent solutions can be exact or approximated.
The algorithm then computes the fluxes through the cell boundaries. 

\subsection{Decoupling}

As the system evolves hydrodynamically, it eventually cools down, and hadronization, known as decoupling, takes place. 
We use the THERMINATOR 2~\cite{THERMINATOR}, a Monte Carlo approach, in the CHESS code to implement the hadrons decoupling using the Cooper-Frye prescriptions~\cite{CooperA, CooperB}. 
The Cooper-Frye formula describes the hadronic decoupling as follows
\begin{equation}
    E\,\frac{d^3 N}{d^3 {\vec{p}}} = \int_{\Sigma} d\Sigma_{\mu}\, \frac{d^3 p}{E}\,p^{\mu}\,f(x, p),
    \label{N_particle-number-full}
\end{equation}
where $d\Sigma_{\mu}$ is an element of the freezeout hypersurface and $f(x,p)$ gives the distribution function, which can be properly modified to take into account the effect of the bulk and shear viscosities.
Besides hadronization, the model also takes into account the two- and three-body decays of resonances.
Although explicit space-time evolution of hadronic degree of freedom will not be considered under the current theoretical framework, we understand that the main results on anisotropy drawn in this study are largely unaffected.

Technically, THERMINATOR 2 provides the freedom to choose whether the freeze-out surface is parameterized in a 2+1 plus boost invariant symmetry or a genuine 3+1 fashion.
In this work, we will adopt a boost invariant scheme for the hydrodynamic evolution and the hadronization.

\subsection{The united code: CHESS}

\begin{figure}[H]
\centerline{\includegraphics[scale=0.8]{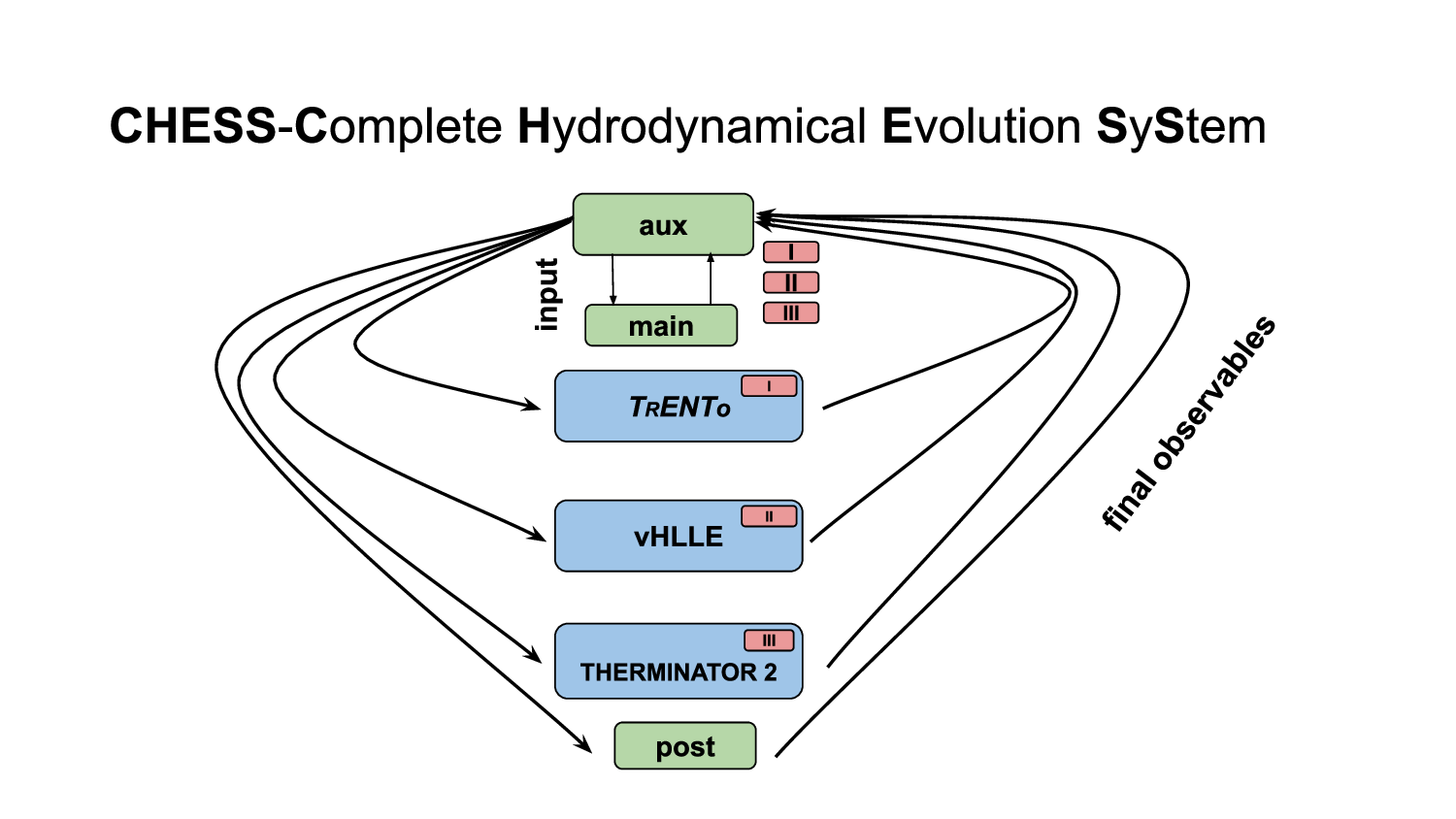}}
\caption{The flowchart for the CHESS code.}
\label{flowchart}
\end{figure}

In the present work, we employ the CHESS code~\cite{chess_ref}, which is a conjunction of the IC implemented by $T_RENT_o$, the relativistic hydrodynamics governed by vHLLE, and the decoupling using the THERMINATOR 2. 
Different code ingredients are united by implementing a series of Python scripts.
The flowchart for the CHESS code is presented in Fig.~\ref{flowchart}.

The green boxes in Fig.~\ref{flowchart} indicate the auxiliary scripts implemented in Python, namely, ``aux'', ``main'', and ``post''.
The blue boxes are the employed numerical packets, and the pink boxes indicate the configurations and parameters for individual codes. 
The ``aux'' scripts invoke the ``main'' responsible for the specific housekeeping tasks.
Among others, it separates the input file into three parts, denoted by I, II, and III. 

These separated files are then returned to ``aux'' and serve as part of the input files for $T_RENT_o$, vHLLE, and THERMINATOR 2. 
The IC is generated by $T_RENT_o$ using the parameters governed by input file I.
Subsequently, vHLLE is invoked for the relativistic hydrodynamical evolution.
The results obtained from $T_RENT_o$ together with the input file part II will be fed to vHLLE as the parameters.
Then, the output from vHLLE will be combined with input file III, which is used as input to THERMINATOR 2 to perform the hadronization.
Via T
Finally, the postscript is utilized to evaluate the relevant observables, such as particle spectrum and collective flow.
The calculations are carried out on an event-by-event fluctuating basis.

\subsection{Model setup and calibration}

This study considers collisions of Gold (Au) and three types of Uranium targets~\cite{Trento} as shown in Tab.~\ref{trento_imple-defor}, where the Uranium nuclei are denoted as U, U2, and U3, respectively.
Besides the original nuclei, we also introduce further modifications to the deformation parameters $\beta_2$ and $\beta_4$ of the Woods-Saxon profile based on Eq.~\eqref{woods-saxon-deformed}.
In particular, we will further consider different geometric deformations for the given nuclei.
By performing a systematic variation of these parameters, we evaluate the resulting collective flow, particularly the elliptic flow $v_2$. 

The remaining parameters, such as $R$ and $a$, are kept unchanged for these calculations.
By varying the initial nucleus' deformation, we aim to explore their impact on the resultant observables, particularly the collective flow. 
The specific Woods-Saxon profile parameters adopted for the present study are given in Tab.~\ref{trento_imple-defor}, which are indicated by an index number enclosed by a pair of angle brackets following the name of the nucleus.
For instance, U$2\langle 3 \rangle$ indicates the third parameter set for the nucleus U2.

\begin{table}[H]
\centering
 \subfloat{
 \centering
 \begin{tabular}{lllll}
\hline
Symbol & $R$ & $a$ & $\beta_2$ & $\beta_4$ \\
\hline
Au & 6.37 & 0.535 & 0.0 & 0.0 \\
U & 6.81 & 0.60 & 0.280 & 0.093 \\
U2 & 6.86 & 0.42 & 0.265 &  0.0 \\
U3 & 6.67 & 0.44 &  0.280 & 0.093 \\
\hline
\end{tabular}}
 \hspace{3cm} 
 \subfloat{
 \centering
 \begin{tabular}{lllll}
\hline
Symbol & $R$ & $a$ & $\beta_2$ & $\beta_4$ \\
\hline
Au$\langle 0\rangle$ & 6.37 & 0.535 & 0.0 & 0.0 \\
Au$\langle 1\rangle$ & 6.37 & 0.535 & -0.065 & -0.031 \\
Au$\langle 2\rangle$ & 6.37 & 0.535 & -0.131 & -0.031 \\
Au$\langle 3\rangle$ & 6.37 & 0.535 & -0.196 & -0.031 \\
Au$\langle 4\rangle$ & 6.37 & 0.535 & -0.262 & -0.031 \\
Au$\langle 5\rangle$ & 6.37 & 0.535 & -0.131 & -0.015 \\
Au$\langle 6\rangle$ & 6.37 & 0.535 & -0.131 & -0.046 \\
Au$\langle 7\rangle$ & 6.37 & 0.535 & -0.131 & -0.062 \\
U$\langle 0\rangle$& 6.81 & 0.60 & 0.0 & 0.0 \\
U$\langle 1\rangle$ & 6.81 & 0.60 & 0.070 & 0.093 \\
U$\langle 2\rangle$ & 6.81 & 0.60 & 0.140 & 0.093 \\
U$\langle 3\rangle$ & 6.81 & 0.60 & 0.210 & 0.093 \\
U$\langle 4\rangle$ & 6.81 & 0.60 & 0.280 & 0.093 \\
U$\langle 5\rangle$ & 6.81 & 0.60 & 0.560 & 0.093 \\
U$\langle 6\rangle$ & 6.81 & 0.60 & 0.280 & 0.046 \\
U$\langle 7\rangle$ & 6.81 & 0.60 & 0.280 & 0.139 \\
U$\langle 8\rangle$ & 6.81 & 0.60 & 0.280 & 0.186 \\
U2$\langle 0\rangle$ & 6.86 & 0.42 & 0.0 & 0.0 \\
U2$\langle 1\rangle$ & 6.86 & 0.42 & 0.265 &  0.0 \\
U3$\langle 0\rangle$ & 6.67 & 0.44 &  0.0 & 0.0 \\
U3$\langle 1\rangle$ & 6.67 & 0.44 & 0.070 & 0.093 \\
U3$\langle 2\rangle$ & 6.67 & 0.44 & 0.140 & 0.093 \\
U3$\langle 3\rangle$ & 6.67 & 0.44 & 0.210 & 0.093 \\
U3$\langle 4\rangle$ & 6.67 & 0.44 & 0.280 & 0.093 \\
U3$\langle 5\rangle$ & 6.67 & 0.44 & 0.560 & 0.093 \\
U3$\langle 6\rangle$ & 6.67 & 0.44 & 0.280 & 0.046 \\
U3$\langle 7\rangle$ & 6.67 & 0.44 & 0.280 & 0.139 \\
U3$\langle 8\rangle$ & 6.67 & 0.44 & 0.280 & 0.186 \\
\hline
\end{tabular}}
\caption{Woods-Saxon profile. Left: Original Au and U nuclei provided by $T_RENT_o$. Right: Implemented Au and U nuclei including the $\beta_2$ and $\beta_4$ variations.}
\label{trento_imple-defor}
\end{table}

\begin{figure}[H]
\centering
\includegraphics[width=0.48\linewidth,height=0.3\linewidth]{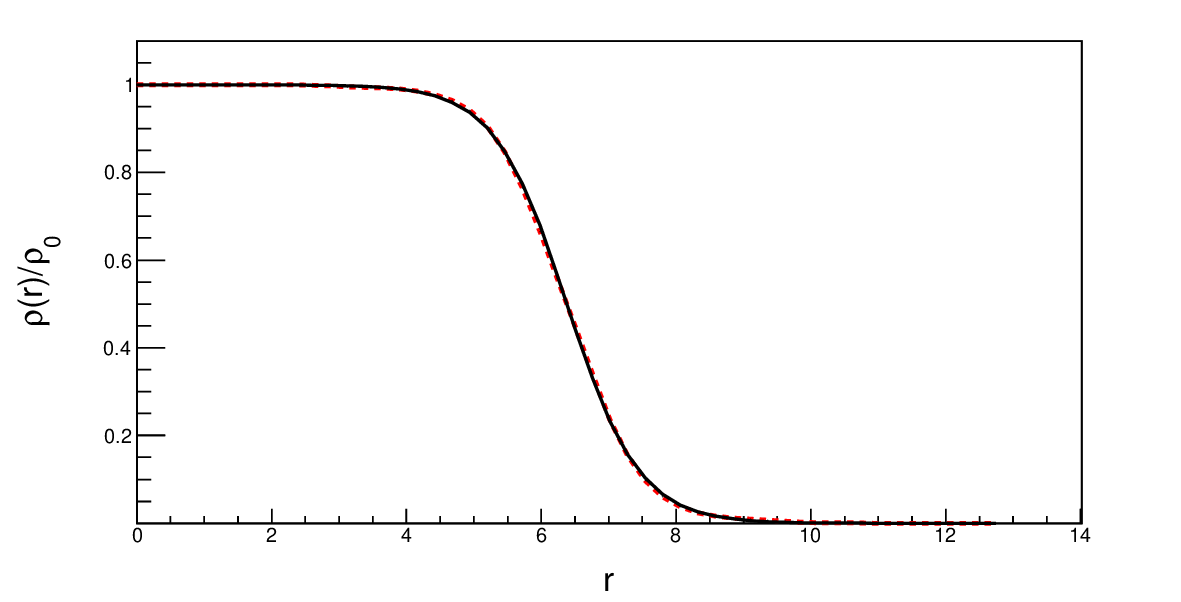}
\includegraphics[width=0.48\linewidth,height=0.3\linewidth]{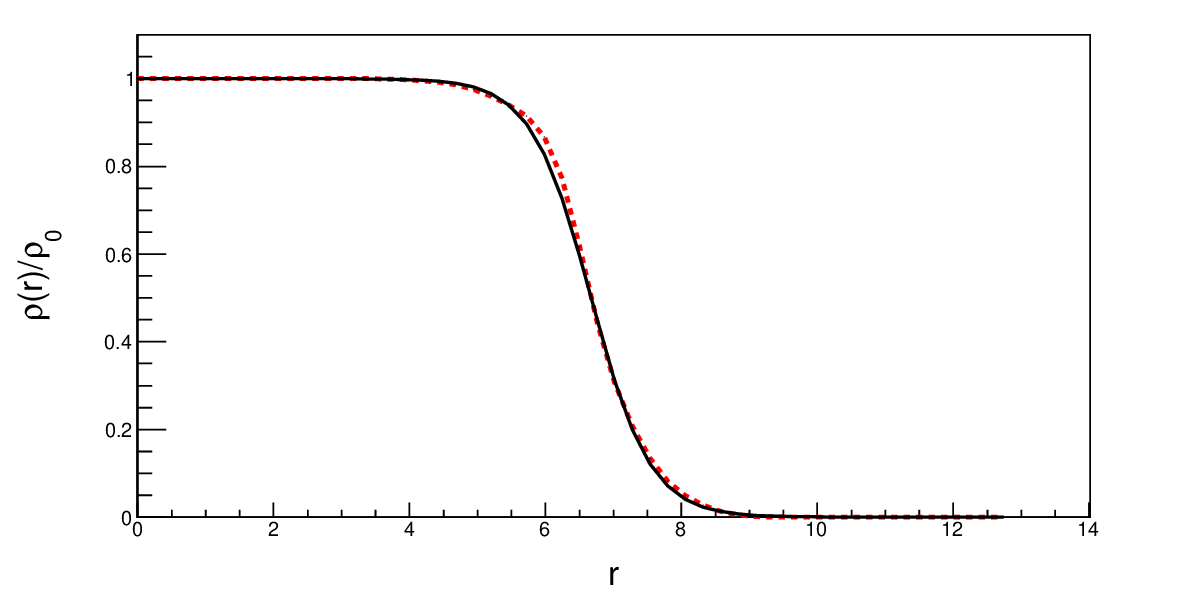}
\includegraphics[width=0.48\linewidth,height=0.3\linewidth]{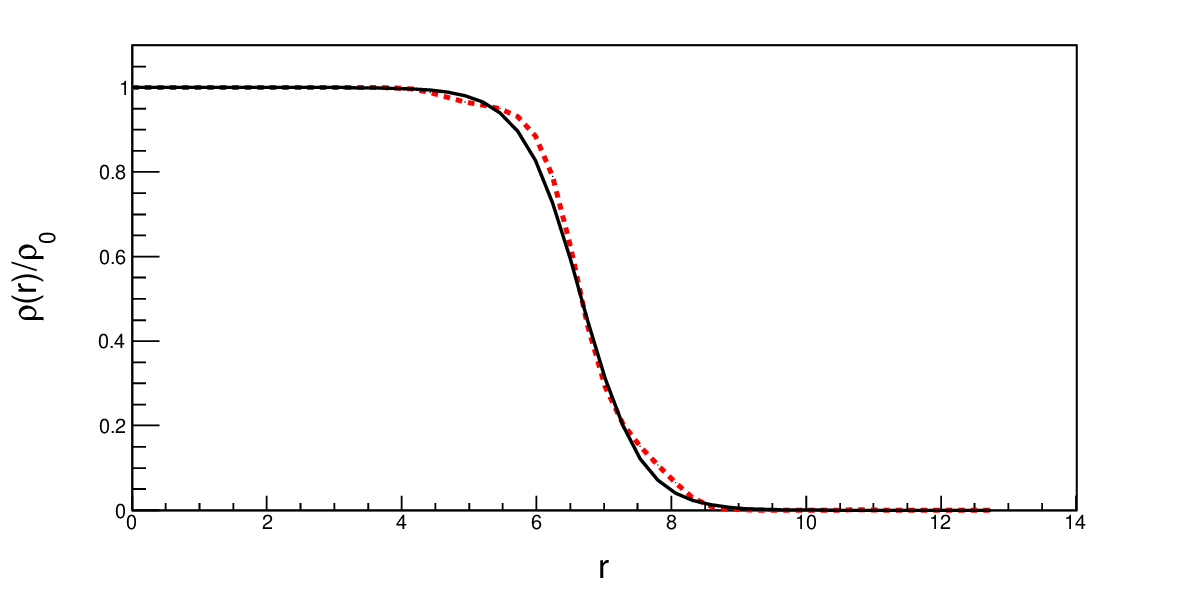}
\caption{Woods-Saxon profile for Au (top left), U2 (top right) and U3 (bottom) nuclei.
The solid black curves are for the symmetric nuclei, while the dashed red ones represent the deformed nuclei.
For the latter, the profile is evaluated along the transverse plane $z=0$, with $\beta_2=-0.131$ and $\beta_4=-0.031$ for Au, and $\beta_2=0.265$ and $\beta_4=0.0$ for U2, and $\beta_2=0.280$ and $\beta_4=0.093$ for U3.}
\label{wds}
\end{figure}

The energy profiles in the radial direction are shown in Fig.~\ref{wds}, where the solid black curves represent the symmetric nuclei, while the dashed red ones are the deformed nuclei. 
One observes that the deviation in the profile is more apparent for the Uranium nucleus, especially if we assume a higher degree of deformation as present at the bottom of Fig.~\ref{wds}.
Visually, however, the deformation does not considerably modify the Woods-Saxon profile.
Nonetheless, as it turns out, the potential impact on the collective flow is substantial regarding observational implications.

We now proceed to calibrate our model to produce the correct multiplicity distribution $dN_{ch}/d\eta$.
The experimental results indicate that the charged particle multiplicity as a function of centrality fraction $x\in [0,1]$ can be parameterized as follows~\cite{star}:
\begin{equation}
    \frac{dN_{ch}}{d\eta} = \left[c_1 - c_2\,x + c_3\,\exp{(-c_4 \,x^{c_5})}\right]^4 , \label{star_parametrization}
\end{equation}
where $c_i$ are constants.
Specifically, we have $c_1=5.0670$, $c_2=3.923$, $c_3=0.2310$, $c_4=18.37$, and $c_5=0.4842$ for Au+Au collisions, and $c_1=5.3473$, $c_2=4.298$, $c_3=0.2959$, $c_4=18.21$, and $c_5=0.4541$ for U+U collisions.
It is noted that the centrality fraction $x$, the inverse of Eq.~\eqref{star_parametrization}, can be estimated from the data since it is essentially defined as a cumulative distribution function as presented in~\cite{Giacalone}:
\begin{equation}
    x\left(\frac{dN_{ch}}{d\eta}\right) 
    \equiv \mathbb{P}\left(y > \frac{dN_{ch}}{d\eta}\right)
    = \int_{{dN_{ch}}/{d\eta}}^{\infty} f(y)dy ,
\end{equation}
where $f(dN_{ch}/d\eta)$ is the probability density function, which is the negative of the reciprocal of Eq.~\eqref{star_parametrization}'s derivative.

\begin{figure}[H]
\centering
\includegraphics[width=0.48\linewidth,height=0.3\linewidth]{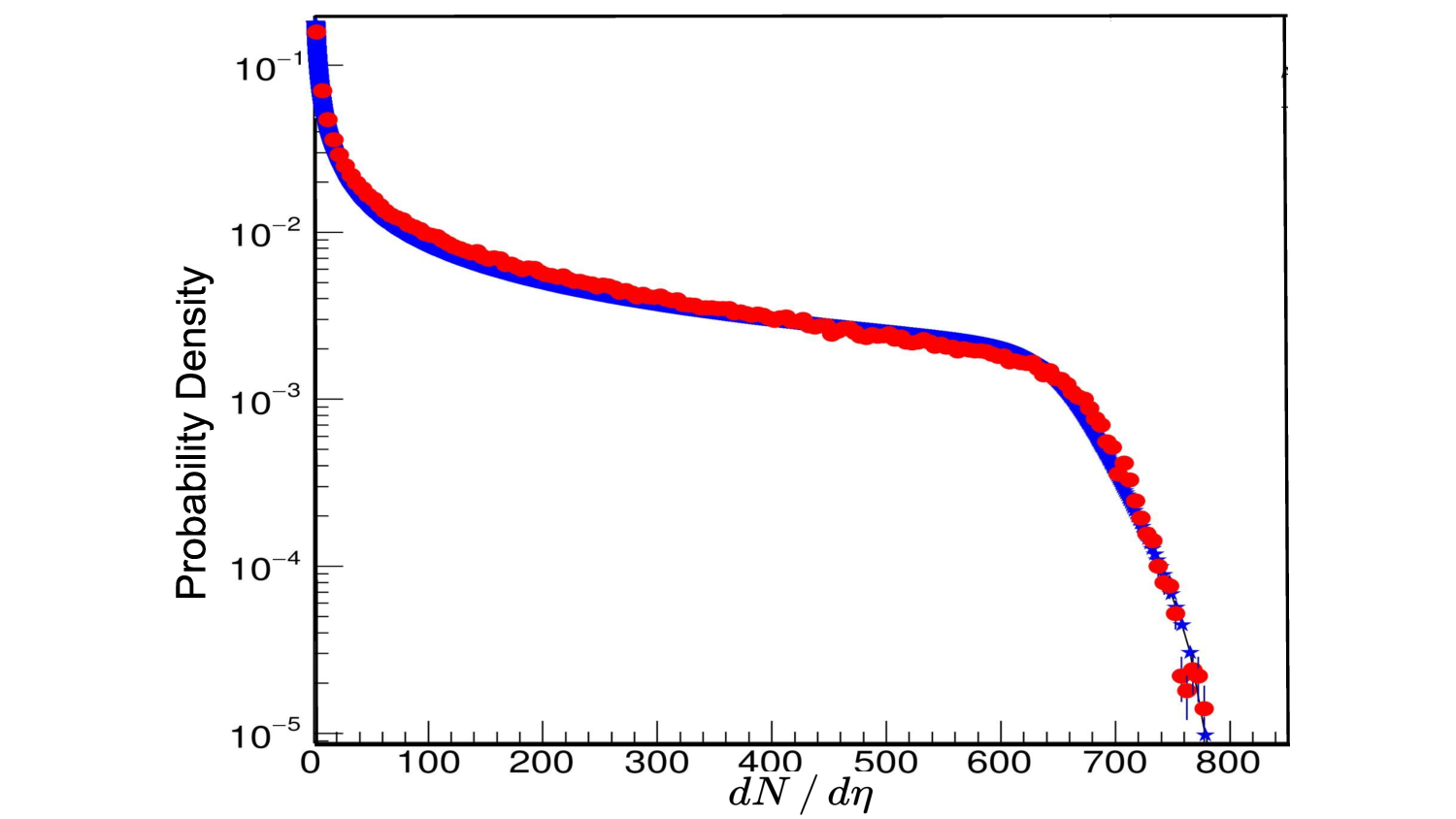}
\includegraphics[width=0.48\linewidth,height=0.3\linewidth]{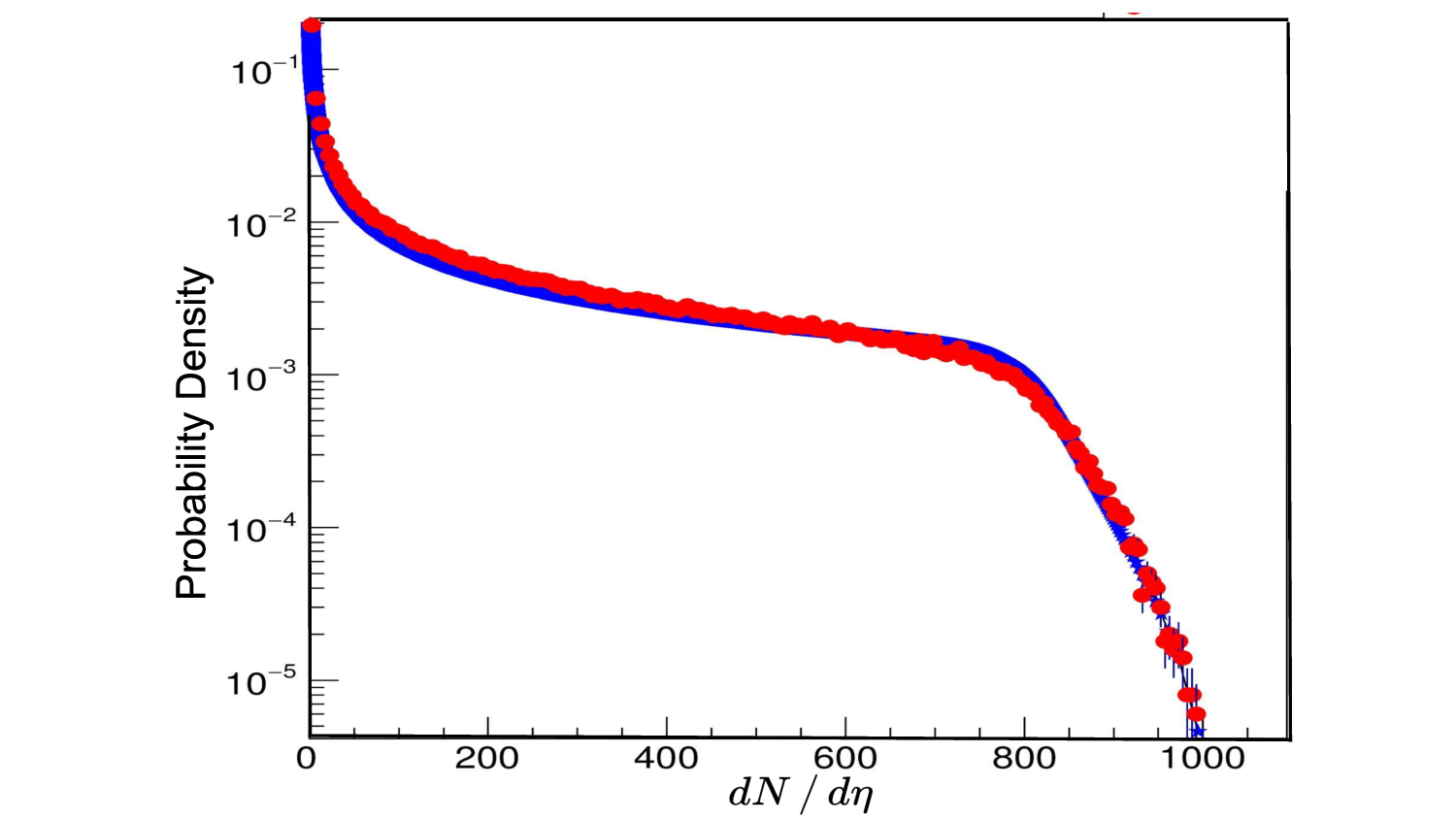}
\caption{Probability density distribution for Au+Au collisions at $\sqrt{s_{NN}}=200$ GeV (left) and U+U ones at $\sqrt{s_{NN}}=193$ GeV (right).
The blue stars represent the results extracted from the STAR data, and the red filled circles are the results using $T_RENT_o$.}
\label{chi_test-Au-U}
\end{figure}

We generate 3200 events to perform the calibration and compare the resultant probability density functions $f$ against those extracted by STAR Collaboration for both the Au+Au and U+U collisions.
Subsequently, the parameters $p, k$, and $w$ of $T_RENT_o$ are tuned to adapt to the experimental data. 
We observe that relevant parameters mostly fall in the following range: $-2\leq p\leq 2; 0 \leq k \leq 2$, and $0.2\leq w \leq 0.6$. 
The parameters of the event generator are obtained by minimizing the sum of the squares of the deviations. 
As shown in Fig.~\ref{chi_test-Au-U}, $T_RENT_o$ code is capable of reproducing the multiplicity yields, where an overall normalization factor is properly introduced in the initial profiles.
The specific values of the parameters are given in Tab.~\ref{trento_parameters}.

\begin{table}[H]
\centering
\begin{tabular}{lllll}
\hline
nucleus & $p$ & $k$ & $w$ & $\chi^2$\\
\hline 
Au &  -0.4 & 1.3  & 0.4 & 0.156\\
U & -0.7 & 1.7 & 0.2 & 0.287\\
\end{tabular}
\caption{The optimized parameters utilized by $T_RENT_o$ event generator.}
\label{trento_parameters}
\end{table}

The nucleus' initial geometry will be implemented in $T_RENT_o$ using the Woods-Saxon profile $\rho(r)$ governed by Eqs~\eqref{woods-saxon} and~\eqref{woods-saxon-deformed}. 
It is noted that in the literature, different profiles have been employed by some authors. 
For example, only the quadrupole expansion has been considered in Refs.~\cite{GiacaloneA, GiacaloneB, GiacaloneC, GiacaloneD}.
On the other hand, in~\cite{JiaA}, quadrupole and octupole deformations have been taken into account for the Woods-Saxon profile in a study for the Zr+Zr and Ru+Ru collisions. 
In the present study, we will consider quadrupole and hexadecapole deformations in terms of the parameters $\beta_2$ and $\beta_4$.

\begin{figure}[H]
\centering
\includegraphics[width=0.48\linewidth,height=0.3\linewidth]{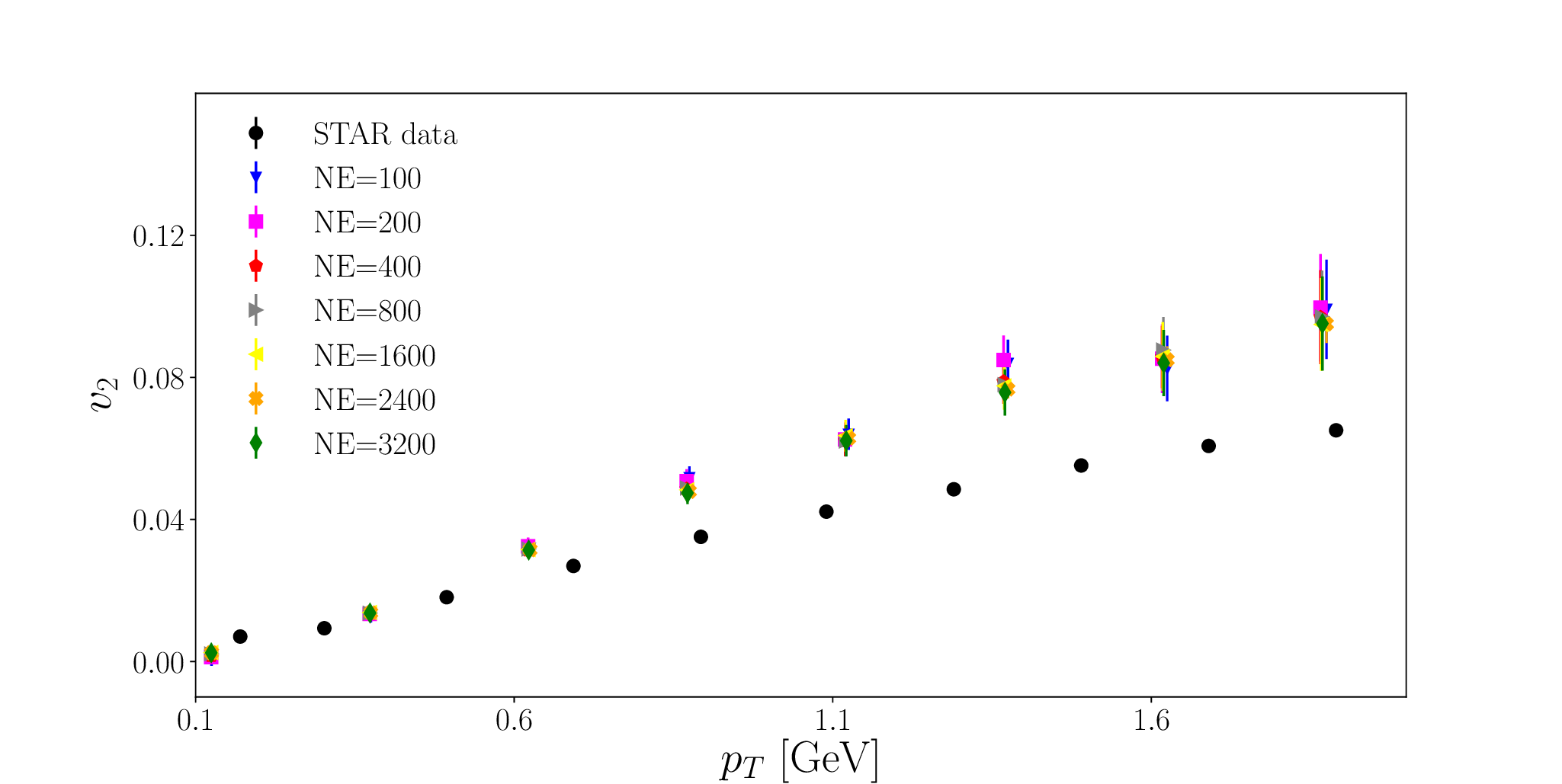}
\includegraphics[width=0.48\linewidth,height=0.3\linewidth]{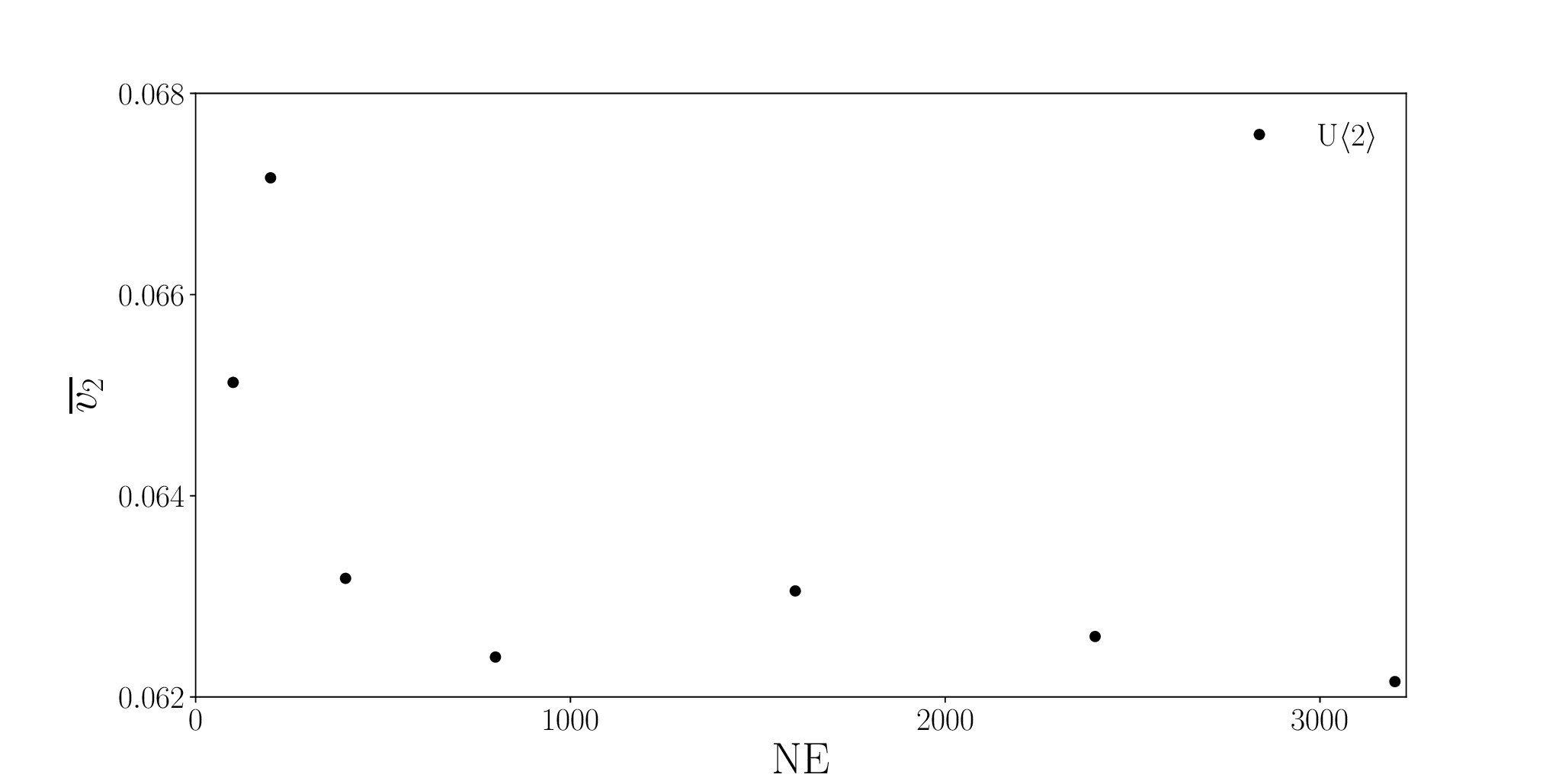}
\caption{Tentative simulations to guarantee the results' statistical power.
Left: The elliptic flow for U$\langle 2\rangle$+U$\langle 2\rangle$ collisions at $\sqrt{s_{NN}}=193$ GeV, evaluated using different NE.
Right: The average elliptic flow $\overline{v_2}$ evaluated as a function of NE.}
\label{comp-Ub}
\end{figure}

Last but not least, to ensure that we have sufficient statistical power for the numerical analysis, tentative simulations are carried out for U$\langle 2\rangle$+U$\langle 2\rangle$ collisions at $\sqrt{s_{NN}}=193$ GeV while varying the number of events (NE).
The flow harmonics evaluated for different NE are presented in Fig.~\ref{comp-Ub}.
In the left panel of Fig.~\ref{comp-Ub}, the differential elliptic flow $v_2$ and its standard error are shown as a function of transverse momentum.
Also, the dependence of average elliptic flow on the NE is analyzed in the right panel of Fig.~\ref{comp-Ub}.

It is observed that both $v_2$ and its fluctuation increase with increasing transverse momentum.
For the calculations of differential flows, reasonable numerical convergence is achieved for NE $\ge 1000$.
Therefore, we adopt the value NE$=1600$ for a given collision setup, which suffices for our purpose.
Also, in our calculations, the number of Monte Carlo is chosen to be 1000.

\section{Numerical results and discussions}\label{section3}

In this section, we carry out numerical simulations aiming to explore the effect of nucleus deformation on the resultant collective flow.
The obtained results on flow harmonics are compared against the experimental data from STAR Collaboration~\cite{Adams2005} for Au+Au collisions at $\sqrt{s_{NN}}=200$ GeV and U+U ones at $\sqrt{s_{NN}}=193$ GeV. 

Using the IC generated by $T_RENT_o$, we performed hydrodynamics simulations using the CHESS code.
There are two factors that involve the IC.
The first one is intrinsic to the nature of the nuclei, namely, the nucleus' radius $R$ and skin depth $a$, particularly for the various Uranium nuclei.
The second factor is the nucleus' initial geometry for different types of nuclei involved in the collisions. 
The present study will primarily focus on the second factor and explore its impact on the observables.

\begin{figure}[H]
\centering
\includegraphics[width=0.48\linewidth,height=0.3\linewidth]{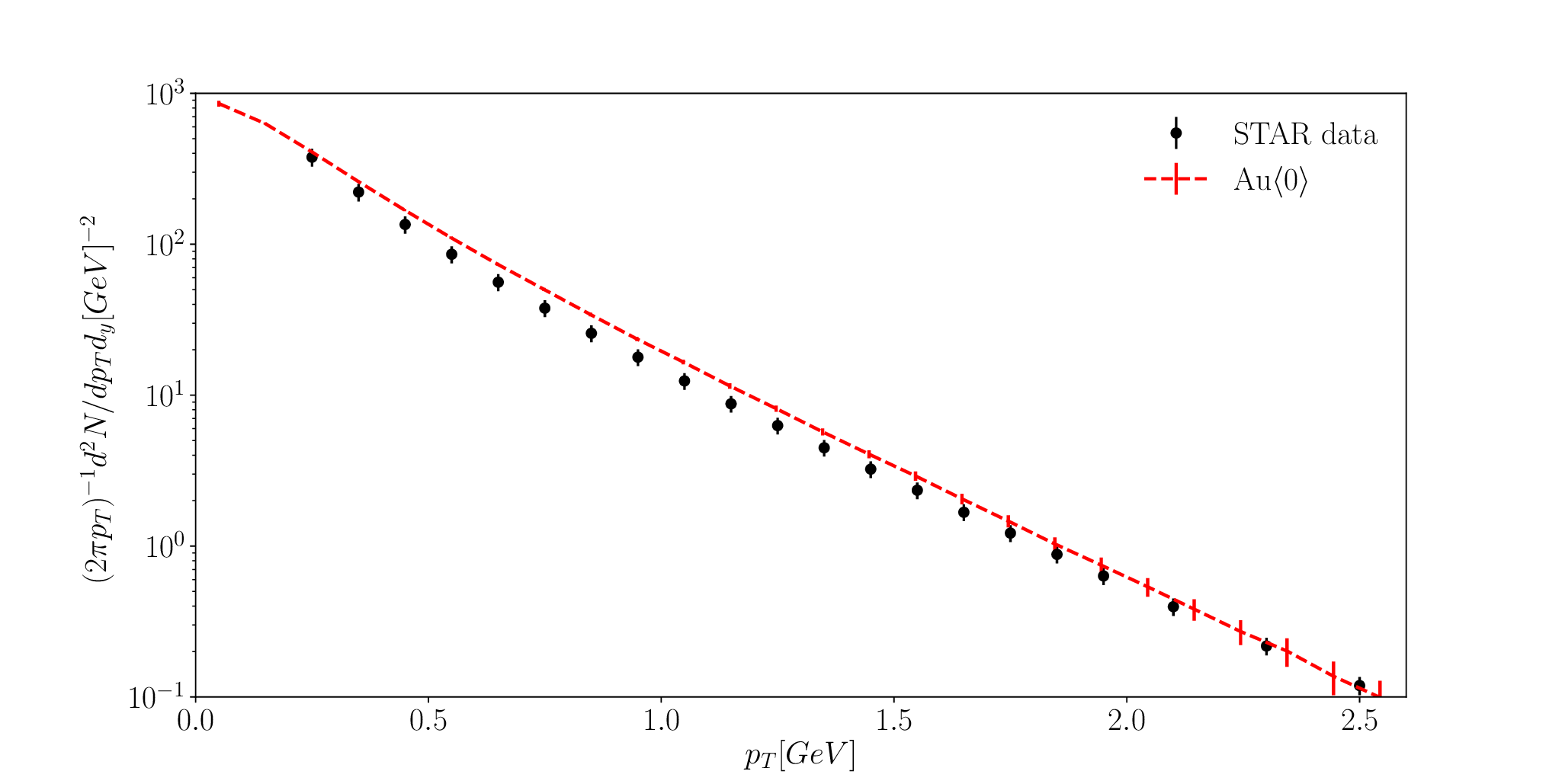}
\includegraphics[width=0.48\linewidth,height=0.3\linewidth]{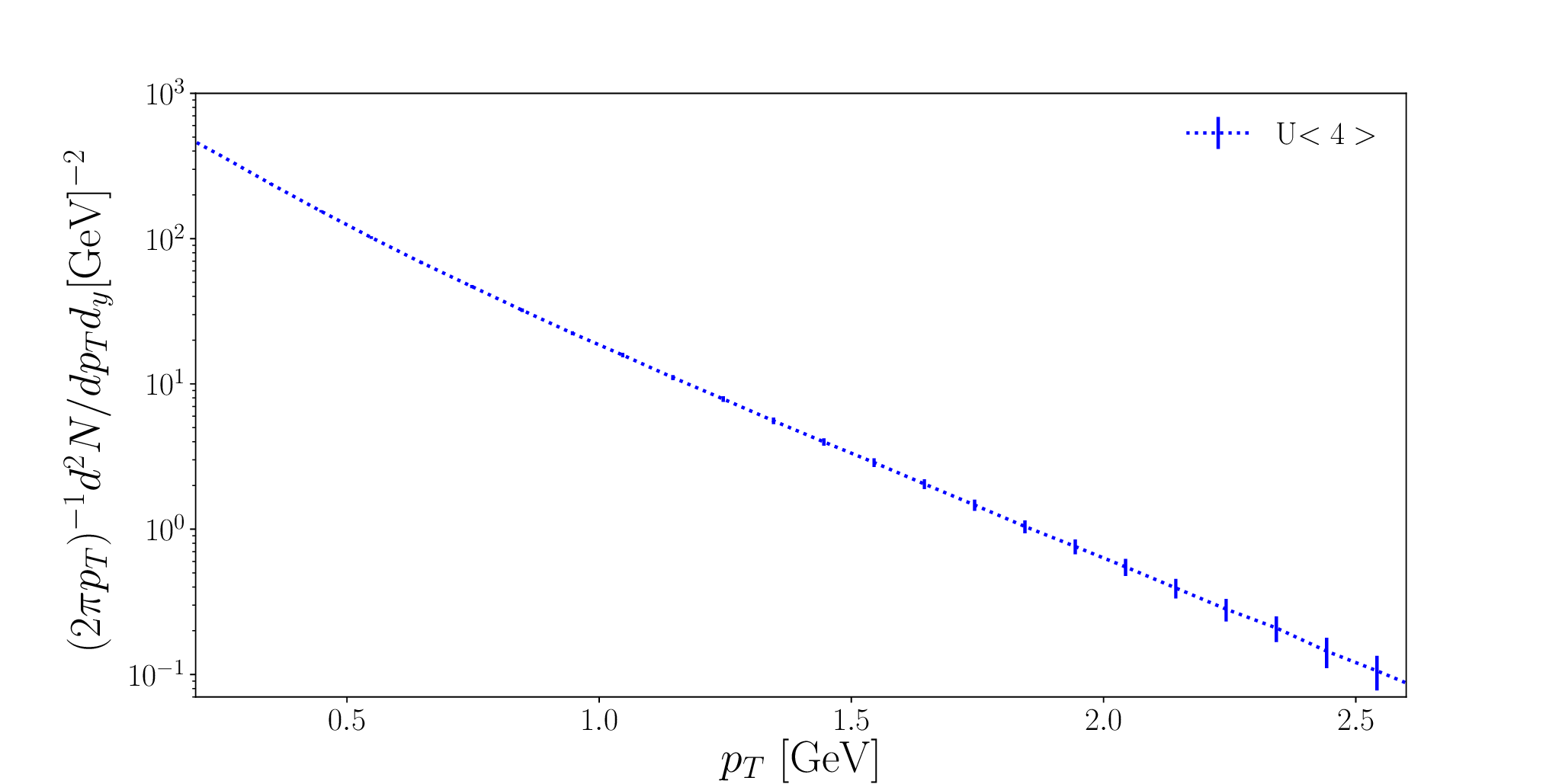}
\caption{Transverse momentum spectra obtained using CHESS.
Left: The numerical results for Au+Au  collisions at $\sqrt{s_{NN}}=200$ GeV (red curve) are compared with the PHOBOS data~\cite{PHOBOS:2002cvz} (black symbols).
Right: The numerical results for U+U collisions at $\sqrt{s_{NN}}=193$ GeV (blue dotted curve).}
\label{Au-multiplicity}
\end{figure}

Before analyzing the flow harmonics, we present the results on the multiplicity spectra.
In Fig.~\ref{Au-multiplicity}, we show the transverse momentum spectra of Au+Au and U+U collisions which are compared with the PHOBOS data~\cite{PHOBOS:2002cvz} when applied for Au+Au collisions at 200 GeV.
The results indicate that reasonable agreement is achieved for both collision systems.
It is noted that the lower transverse momentum region $p_T < 0.5 GeV$ is more relevant for us.
The significance of the emanated lower $p_T$ particles is related to not only the overall multiplicity but also the collective flow, which is primarily governed by the momentum anisotropy transformed from the initial state deformation and fluctuations via the hydrodynamic evolution. 

Regarding the collective flow, we first explore the elliptic flow $v_2$ and its dependence on the initial geometric deformation. 
It was pointed out~\cite{Giacalone} that initial-state anisotropy $\epsilon_2$ affects the final-state elliptic flow through a largely linear relation
\begin{equation}
    v_2\{2\} = \kappa_2\,\epsilon_2\{2\} ,    \label{ani-final-flow}
\end{equation}
where $v_2\{2\}$ is the elliptic flow evaluated using the two-particle cumulant, $\epsilon_2\{2\}$ is the eccentricity second order cumulant, and $\kappa_2$ is a constant.
However, for higher-order harmonics, such a relation no longer holds~\cite{sph-corr-30}.

In terms of quadrupole and hexadecapole deformations, it is unclear how much initial geometric deformation of an individual nucleus will affect the eccentricities and the resulting collective flows on an event-by-event basis.
In this regard, it is interesting to study both the overall deformation and individual effects associated with quadrupole and hexadecapole deformations.

\begin{figure}[H]
\centering
\includegraphics[width=0.48\linewidth,height=0.3\linewidth]{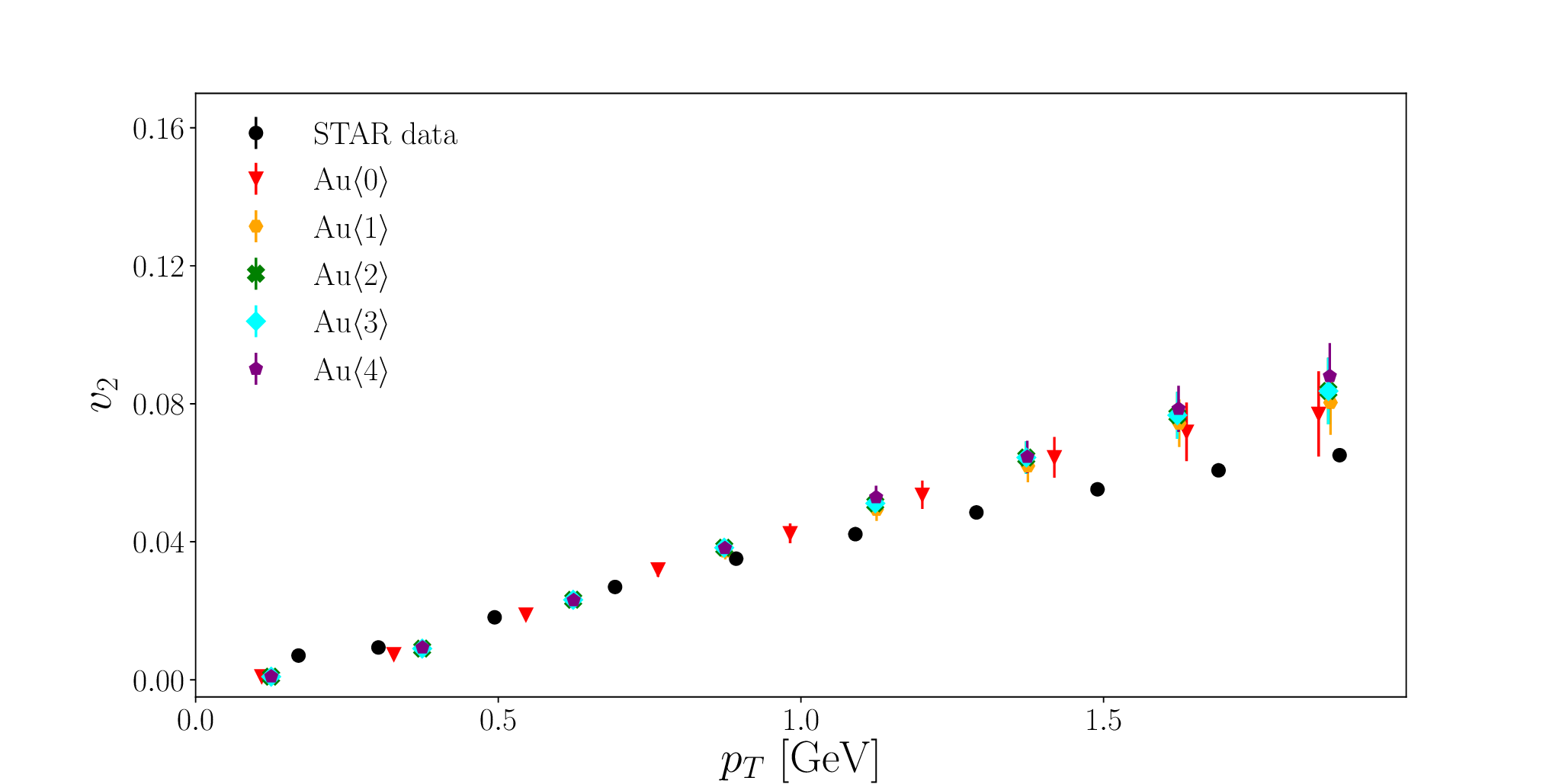}
\includegraphics[width=0.48\linewidth,height=0.3\linewidth]{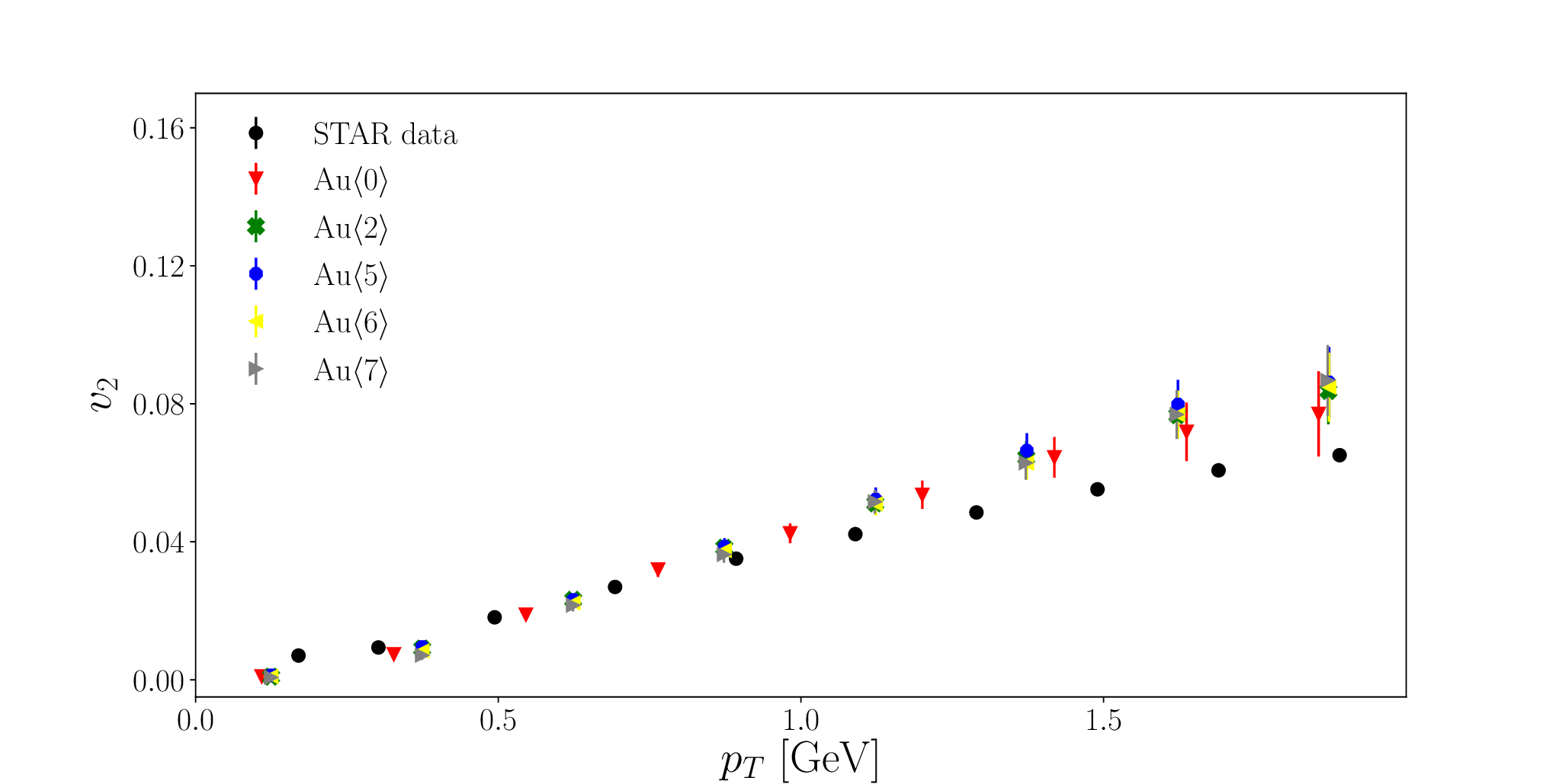}
\includegraphics[width=0.48\linewidth,height=0.3\linewidth]{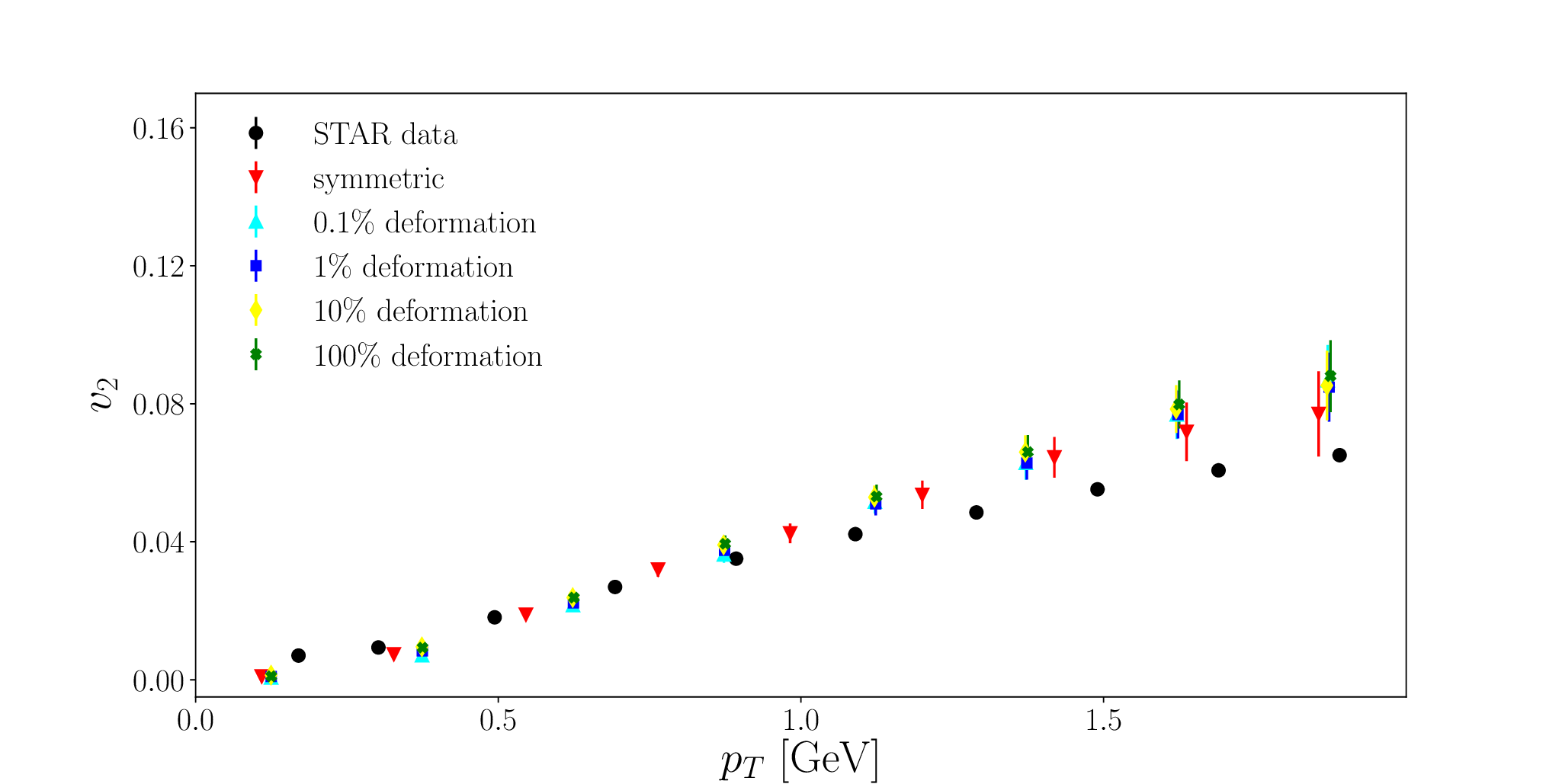}
\caption{Differential elliptic flow $v_2$ using the centrality 0-6\% for Au+Au collisions at $\sqrt{s_{NN}}=200$ GeV with different degrees of nucleus deformation.
The numerical results obtained by the CHESS code are compared with the STAR data~\cite{Adams2005}.
Top-Left: The results obtained by varying quadrupole deformation parameter $\beta_2$, the specific parameters are given in Tab.~\ref{trento_imple-defor}.
Top-Right: The results obtained by varying hexadecapole deformation parameter $\beta_4$, the specific parameters are given in Tab.~\ref{trento_imple-defor}.
Bottom: The results obtained by varying the overall deformations of the Au nucleus.
For all the plots, the red triangles are used to represent the results for symmetric Au nuclei.
The error bars indicate the standard errors.}
\label{def-perc}
\end{figure}

In Fig.~\ref{def-perc}, we show the elliptic flows for Au+Au collisions by varying $\beta_2$ and $\beta_4$ while keeping the remaining parameters unchanged.
Moreover, we elaborate on a scenario when one gradually increases the overall deformation of the nucleus.
Specifically, for the latter case, the symmetric nucleus ($0\%$), $0.1\%$, $1\%$, $10\%$, and $100\%$ percent of deformation are considered. 
The relative deformations are implemented as an interpolation between the symmetric and Au<2> nuclei and carried out simultaneously for $\beta_2$ and $\beta_4$.
The resulting elliptic flows are presented as functions of $p_T$ and are compared with the STAR data~\cite{Adams2005}.

It is observed that the initial geometrical deformations do affect the resulting elliptic flow.
As shown in the bottom panel of Fig.~\ref{def-perc}, the resulting differential flow slightly increases as the overall deformation increases.
The relation between the size of the deformation and the resulting impact on the elliptic flow is primarily linear. 
This is understood regarding existing results, which show that the elliptic flow $v_2$ is mainly proportional to the second cumulant of initial space anisotropy $\epsilon_2$~\cite{hydro-vn-34, sph-vn-04, hydro-vn-45, hydro-vn-55, sph-corr-30}, except that the elliptic flow persists for ultra-central collisions~\cite{LHC-cms-vn-05, hydro-vn-ph-20, hydro-vn-55, hydro-ml-bayesian-23}. 
Nonetheless, the model could not faithfully reproduce the data at large transverse momenta.
By tuning individual deformation parameters, the above results mostly hold.
Specifically, the elliptic flow generally becomes more suppressed, particularly at the region with larger transverse momenta, as $\beta_2$ or $\beta_4$ increases.
At smaller $p_T$, on the other hand, the dependence of elliptic flow on these parameters is complex.
An explicit study on the dependence of $v_2$ on the deformation parameters will be presented below in Fig.~\ref{au-av2-b2}, where the average flow will be analyzed.

\begin{figure}[H]
\centering
\includegraphics[width=0.48\linewidth,height=0.3\linewidth]{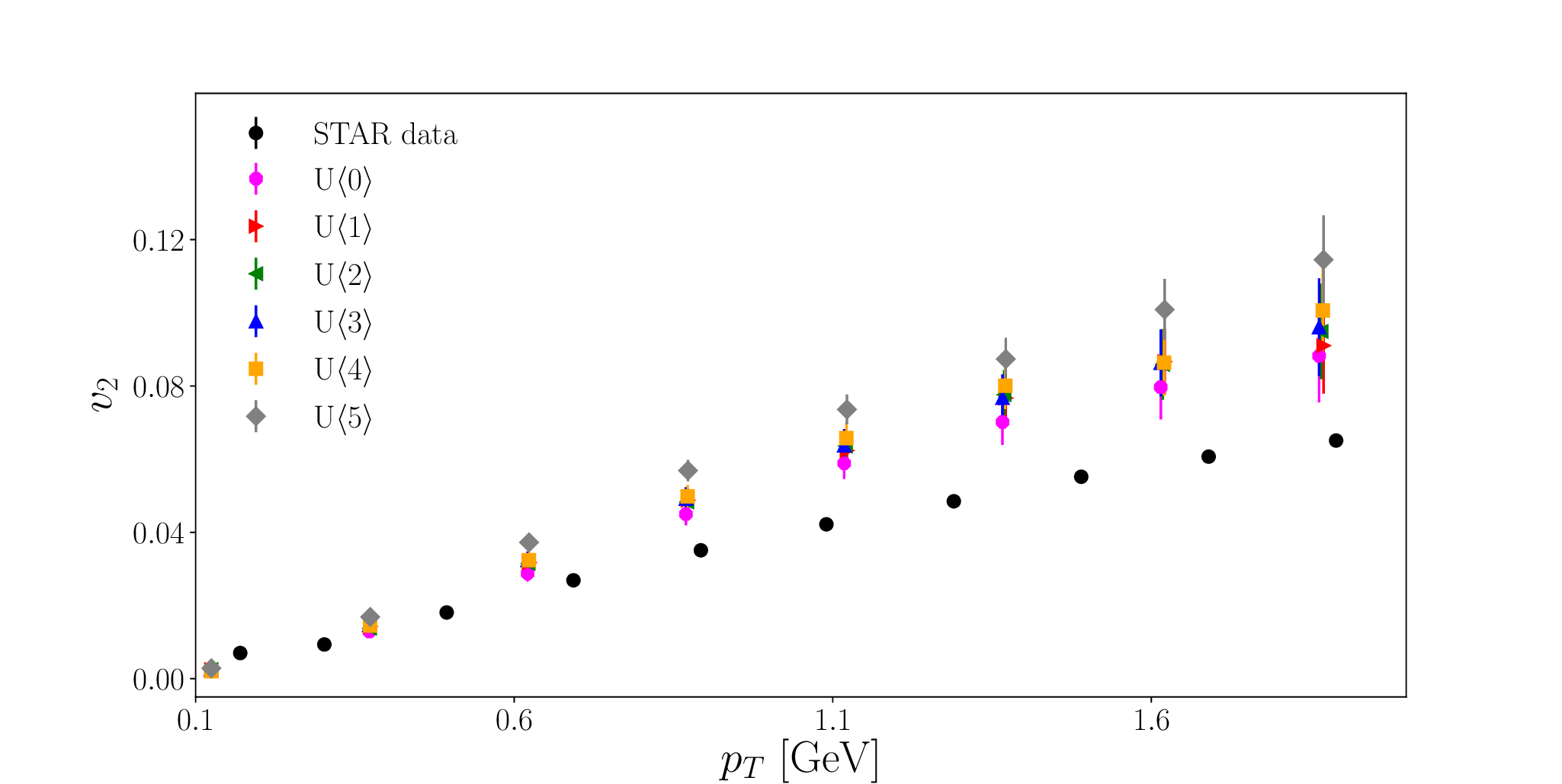}
\includegraphics[width=0.48\linewidth,height=0.3\linewidth]{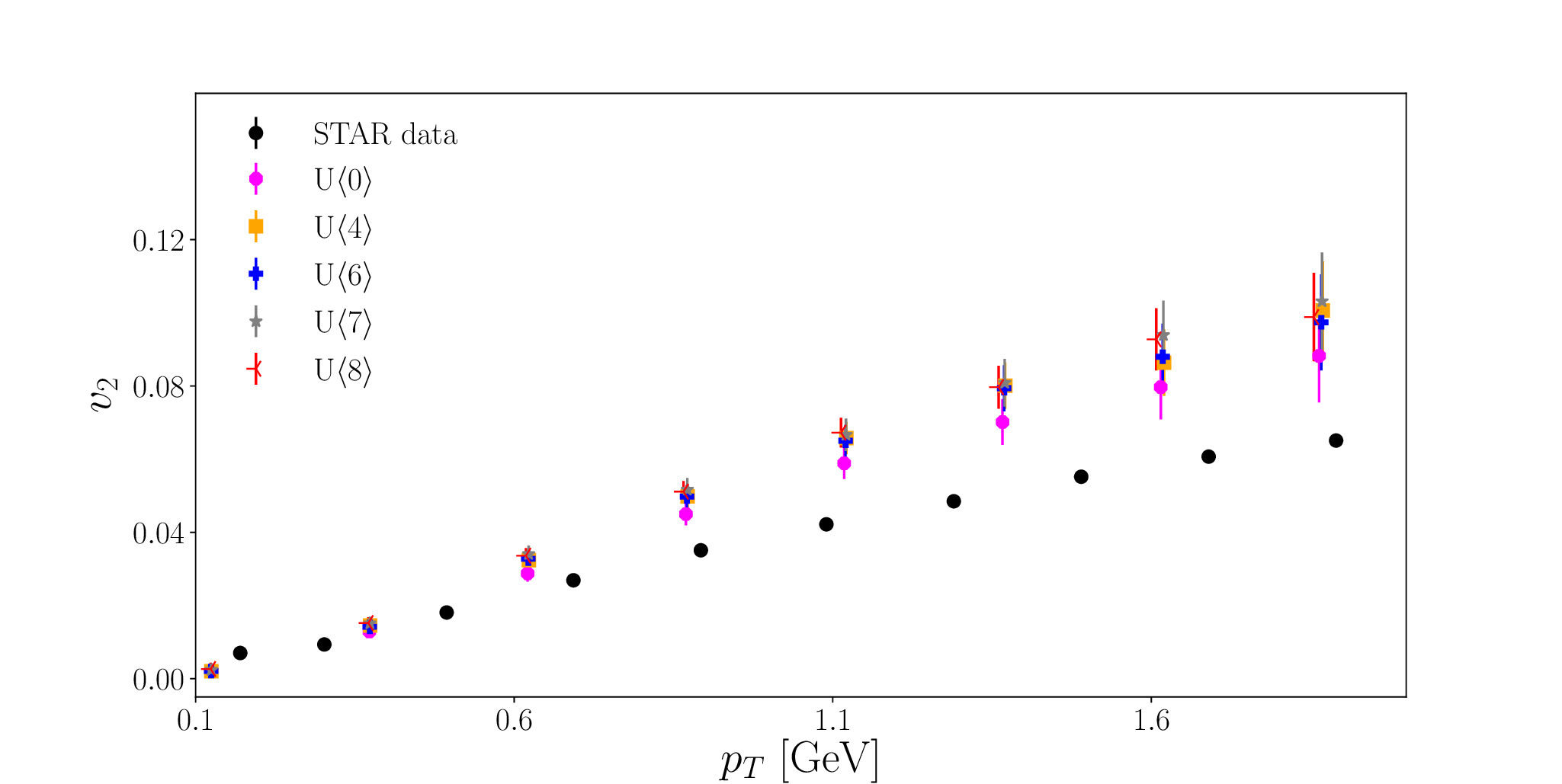}
\includegraphics[width=0.48\linewidth,height=0.3\linewidth]{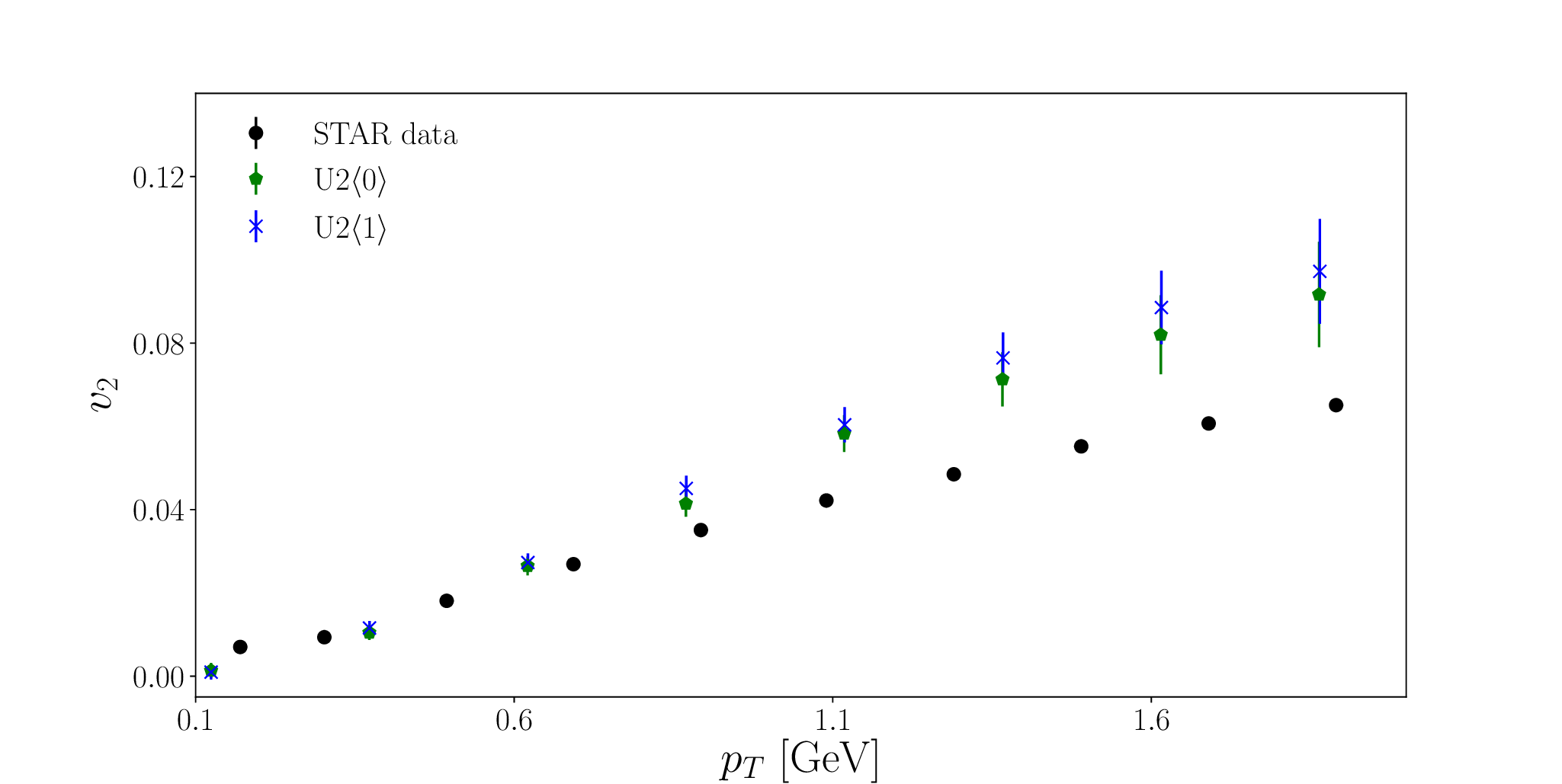}
\includegraphics[width=0.48\linewidth,height=0.3\linewidth]{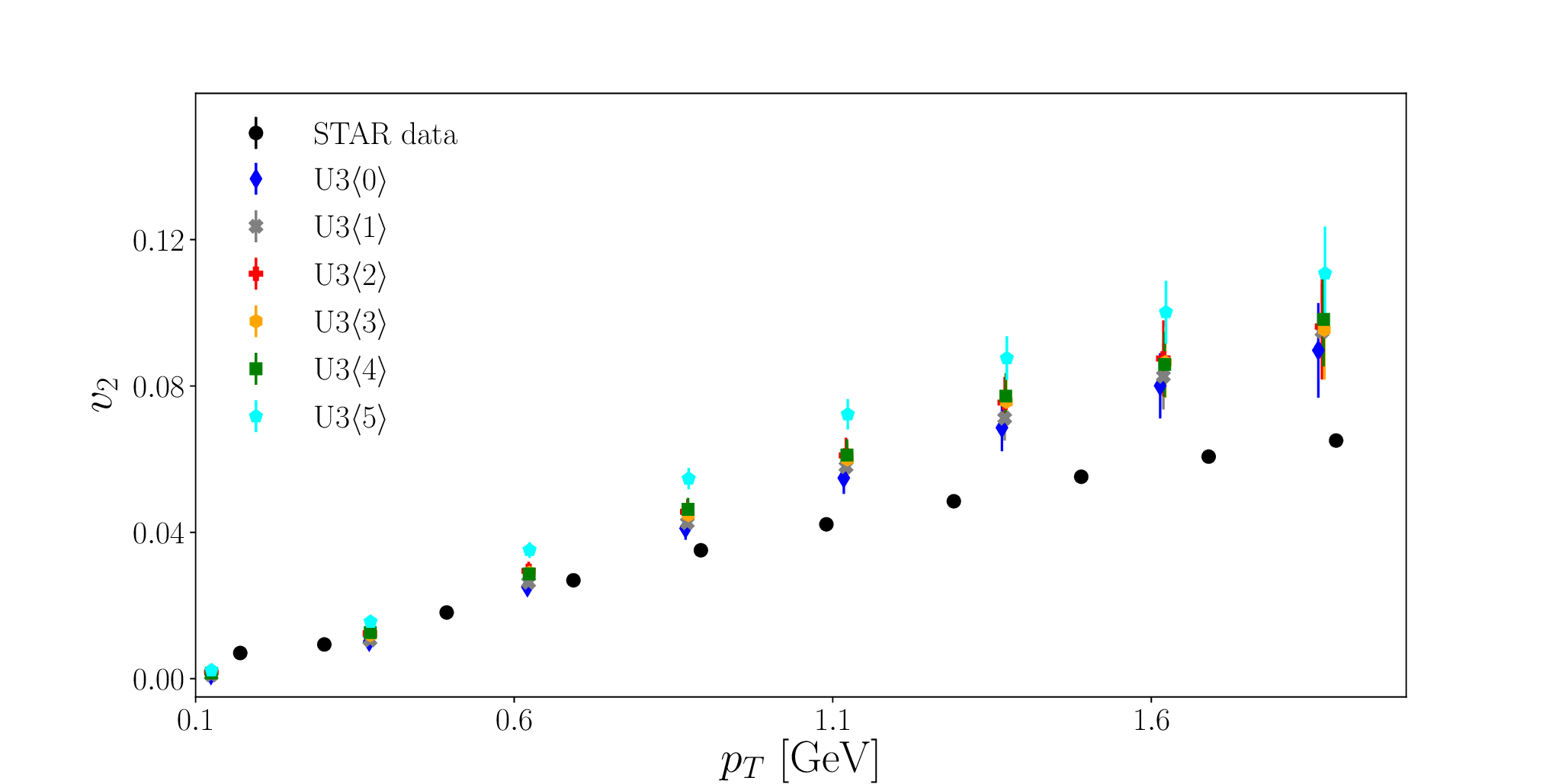}
\includegraphics[width=0.48\linewidth,height=0.3\linewidth]{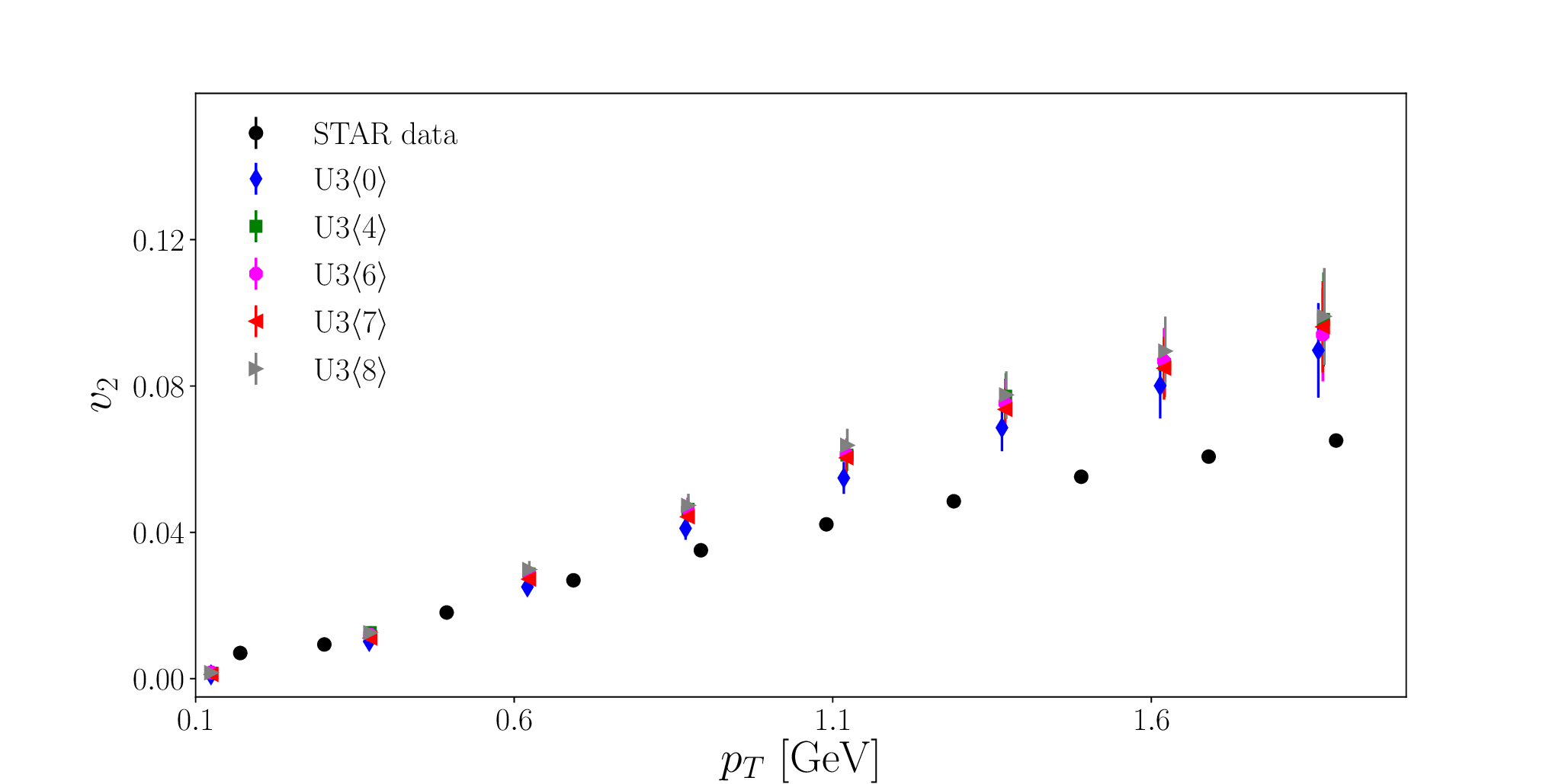}
\includegraphics[width=0.48\linewidth,height=0.3\linewidth]{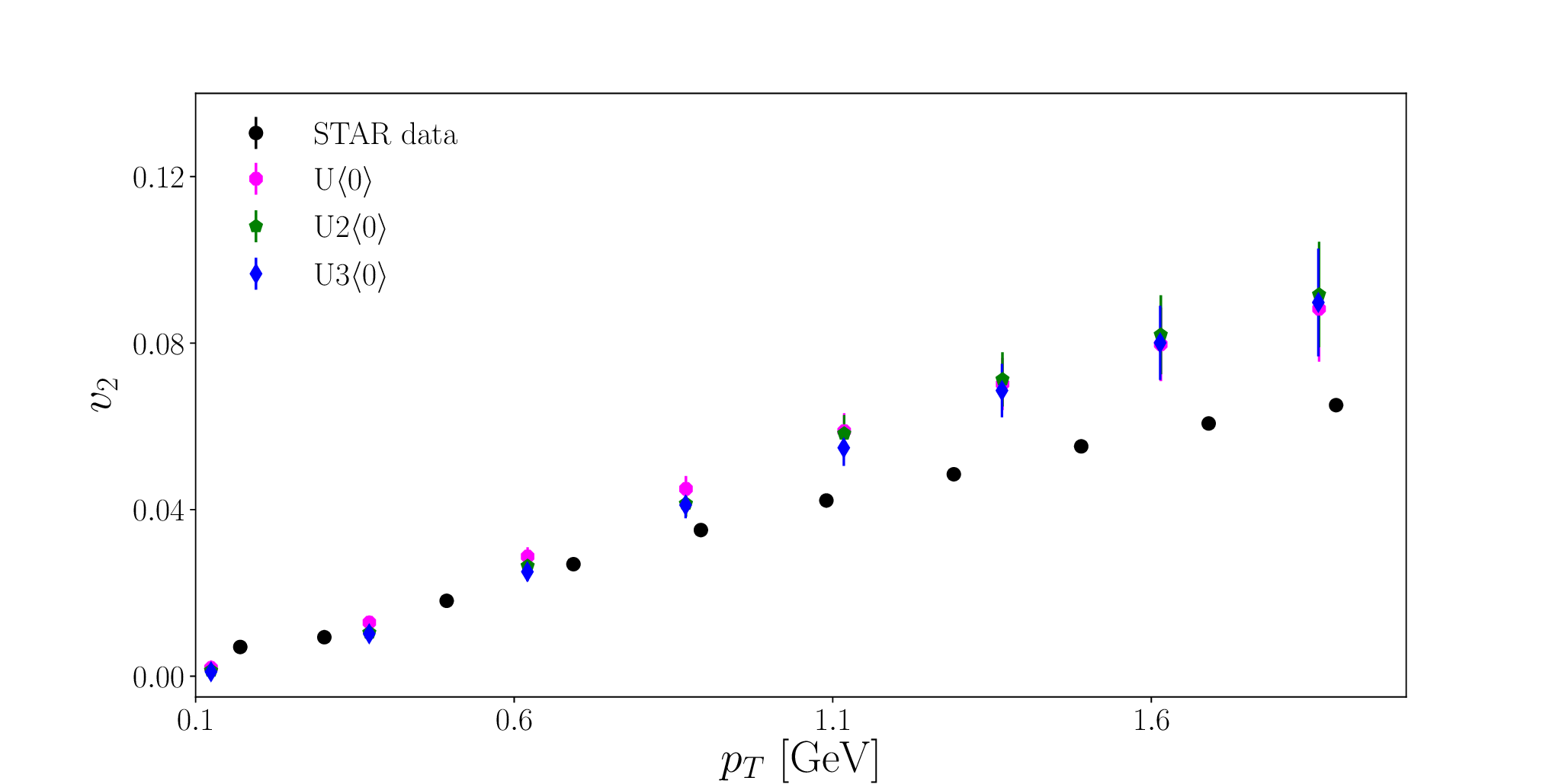}
\caption{Differential elliptic flow $v_2$ for U+U collisions at $\sqrt{s_{NN}}=193$ GeV with centrality 0-6\% and different degrees of nucleus deformation.
The numerical results obtained by the CHESS code are compared with the STAR data~\cite{Adams2005} for Au+Au collisions at 200 GeV.
Top-Left: The results obtained by varying quadrupole deformation parameter $\beta_2$ for U.
Top-Right: The results obtained by varying hexadecapole deformation parameter $\beta_4$ for U.
Middle-Left: The results obtained by varying quadrupole deformation parameter for U2.
Middle-Right: The results obtained by varying hexadecapole deformation parameter $\beta_2$ for U3.
Bottom-left: The results obtained by varying hexadecapole deformation parameter $\beta_4$ for U3.
Bottom-right: The results obtained for symmetric U+U, U2+U2, and U3+U3 collisions.
The specific parameters are given in Tab.~\ref{trento_imple-defor} for all the plots.
The error bars indicate the standard errors.}
\label{U-beta-full-b2}
\end{figure}

A similar procedure is also carried out for various types of U+U collisions.
In Fig.~\ref{U-beta-full-b2}, we present the flow harmonics for the three Uranium nuclei while tuning the respective parameters regarding the quadrupole and hexadecapole deformations.
The numerical results obtained by CHESS are also compared to the STAR data~\cite{Adams2005}. 
Generally, the initial geometric deformation plays a role in the final elliptic flow.
As expected, the differential elliptic flow mainly increases with increasing $\beta_2$ or $\beta_4$, which is apparent, particularly for the region with large transverse momentum.

This feature is found to be universal for all three types of U+U collisions.
On the other hand, at the region with insignificant $p_T$, the dependence of $v_2$ on the deformation parameter is more subtle, which will be explored further below.
Regarding the magnitude of the effect, one concludes that hexadecapole deformation has a less significant impact when compared to that of quadrupole deformation.
Even though $\beta_2$ does not affect the initial eccentricity $\epsilon_2$ in a straightforward fashion, it is understood that $\beta_4$ is likely to have a minor effect compared to $\beta_2$.
The obtained results agree reasonably with the STAR data for the low $p_T$ region.

However, deviations occur at larger transverse momenta, where the numerical simulations systematically overshot the data.
In the bottom-right panel of Fig.~\ref{U-beta-full-b2}, we also perform a comparison for the three types of uranium nuclei when eliminating their geometric deformations.
It is found that the resultant elliptic flows are somewhat similar for the three systems, while U$\langle 0\rangle$+U$\langle 0\rangle$ collisions turn out to stay slightly closer to the experimental data.

By comparing the results of Au+Au and U+U collisions, one seems to draw somewhat different observations about the effect of deformation on the collective flow.
On the one hand, for U+U collisions, we have observed that the increase of deformation has, by and large, promoted the elliptic flow $v_2$, in accordance with Eq.~\eqref{ani-final-flow}.
On the other hand, for Au+Au collisions, the observable effect is less substantial.
In particular, at the high transverse momentum region, while the numerical results overshot the data, flow harmonics only increase slightly as the deformation becomes more significant.
In other words, one finds that the elliptic flow in U+U collisions is more sensitive to the deformation.
We attribute the difference in the magnitude of the effect to the highly nonlinear nature of the underlying hydrodynamic evolution and the strongly interacting system's equation of state.

\begin{table}[H]
\centering
\begin{tabular}{llll}
\hline
nucleus & $b$ & $N_\mathrm{part}$ & $\epsilon_2$\\
\hline
Au$\langle 0\rangle$ & 9.82 & 104 & 0.432\\
Au$\langle 3\rangle$ & 9.90 & 104 & 0.429\\
Au$\langle 7\rangle$ & 10.04 & 99 & 0.446\\
U$\langle 0\rangle$ & 10.69 & 121 & 0.489\\
U$\langle 4\rangle$ & 10.94 & 118 & 0.498\\
U$\langle 8\rangle$ & 10.81 & 118 & 0.495\\
U2$\langle 0\rangle$ & 10.09 & 123 & 0.513\\ 
U2$\langle 1\rangle$ & 10.28 & 121 & 0.525\\ 
U3$\langle 0\rangle$ &  9.86 & 127 & 0.503\\
U3$\langle 4\rangle$ & 10.05 & 126 & 0.509\\
U3$\langle 8\rangle$ & 9.99 & 124 & 0.505\\
\hline
\end{tabular}
\caption{The IC generated by $T_RENT_o$ for the collisions studied in this study.
For specific types of nuclei, the information on impact parameter $b$, the number of participants $N_\mathrm{part}$, and spatial eccentricity $\epsilon_2$ are presented.
When it applies, the table gives average values evaluated on an event-by-event basis.}
\label{initial-conditions-trento}
\end{table}

It is also interesting to validate the relation given by Eq.~\eqref{ani-final-flow}, particularly regarding the apparently opposite effects of the deformation.
To this end, we present the IC, inclusively the initial geometric deformation, generated by $T_RENT_o$ in Tab.~\ref{initial-conditions-trento} on an event-by-event basis.

From Tab.~\ref{initial-conditions-trento}, the observed results on flow and nucleus deformation can be readily understood in terms of Eq.~\eqref{ani-final-flow} as a result of the system's initial geometry.
Specifically, the eccentricity of the IC increases while the impact parameter increases and the number of participants decreases.
As a result, elliptic flow also deviates from the value of symmetric nuclei in alignment with the initial state eccentricity.
These results are largely consistent with the linear response dictated by Eq.~\eqref{ani-final-flow}~\cite{hydro-vn-34, sph-vn-04, sph-corr-30}.
We also have observed that the effect of deformation factor $\beta_4$ seems to play a less significant role in the initial state and final observables than $\beta_2$ observing directly Tab.~\ref{initial-conditions-trento}. 

We compare the elliptic flow data from symmetric Au+Au collisions with symmetric and deformed U+U data, the results can be visualized in Fig.~\ref{Au-U-comp}:

\begin{figure}[H]
\centering
\includegraphics[width=0.48\linewidth,height=0.3\linewidth]{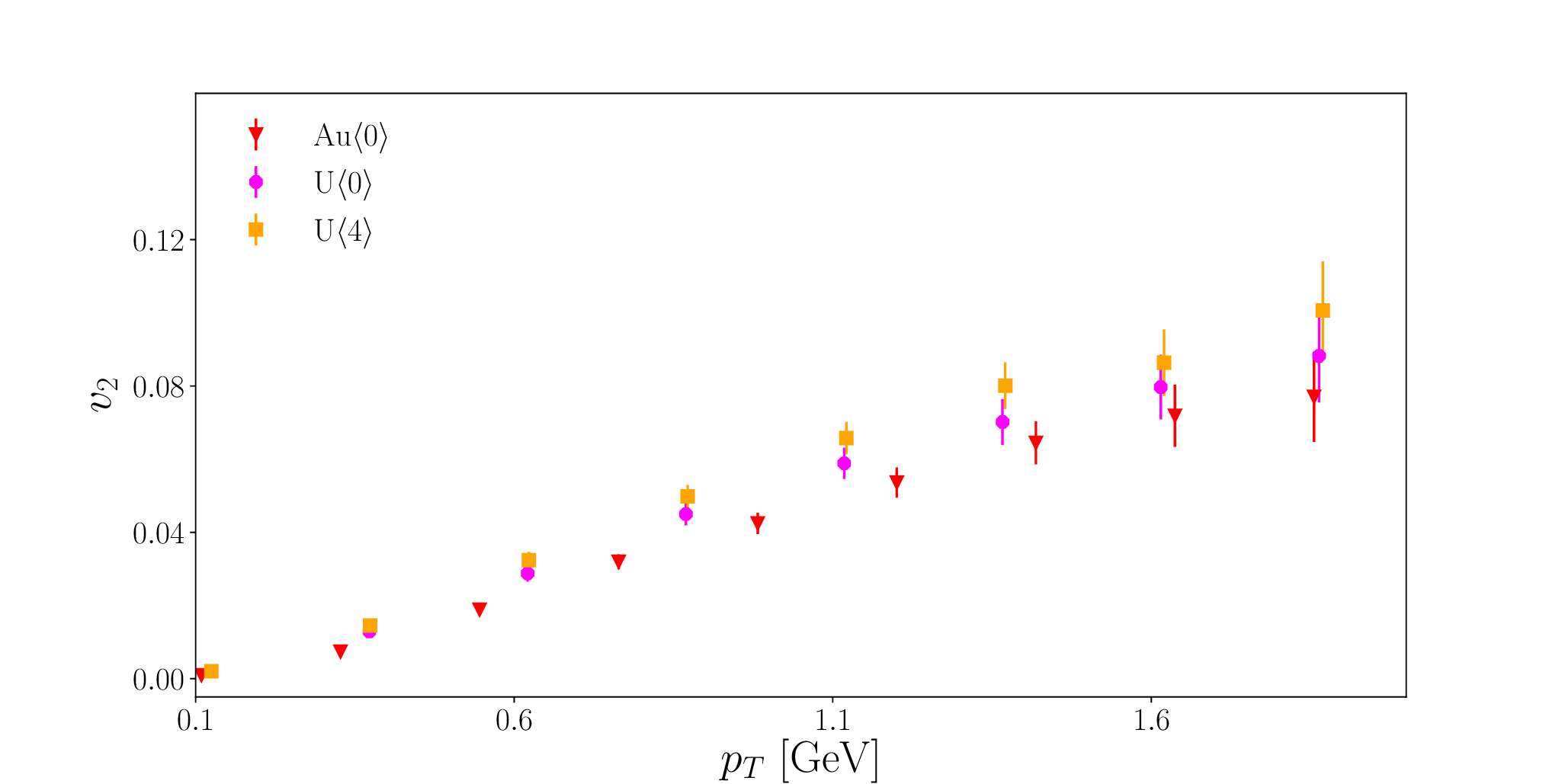}
\includegraphics[width=0.48\linewidth,height=0.3\linewidth]{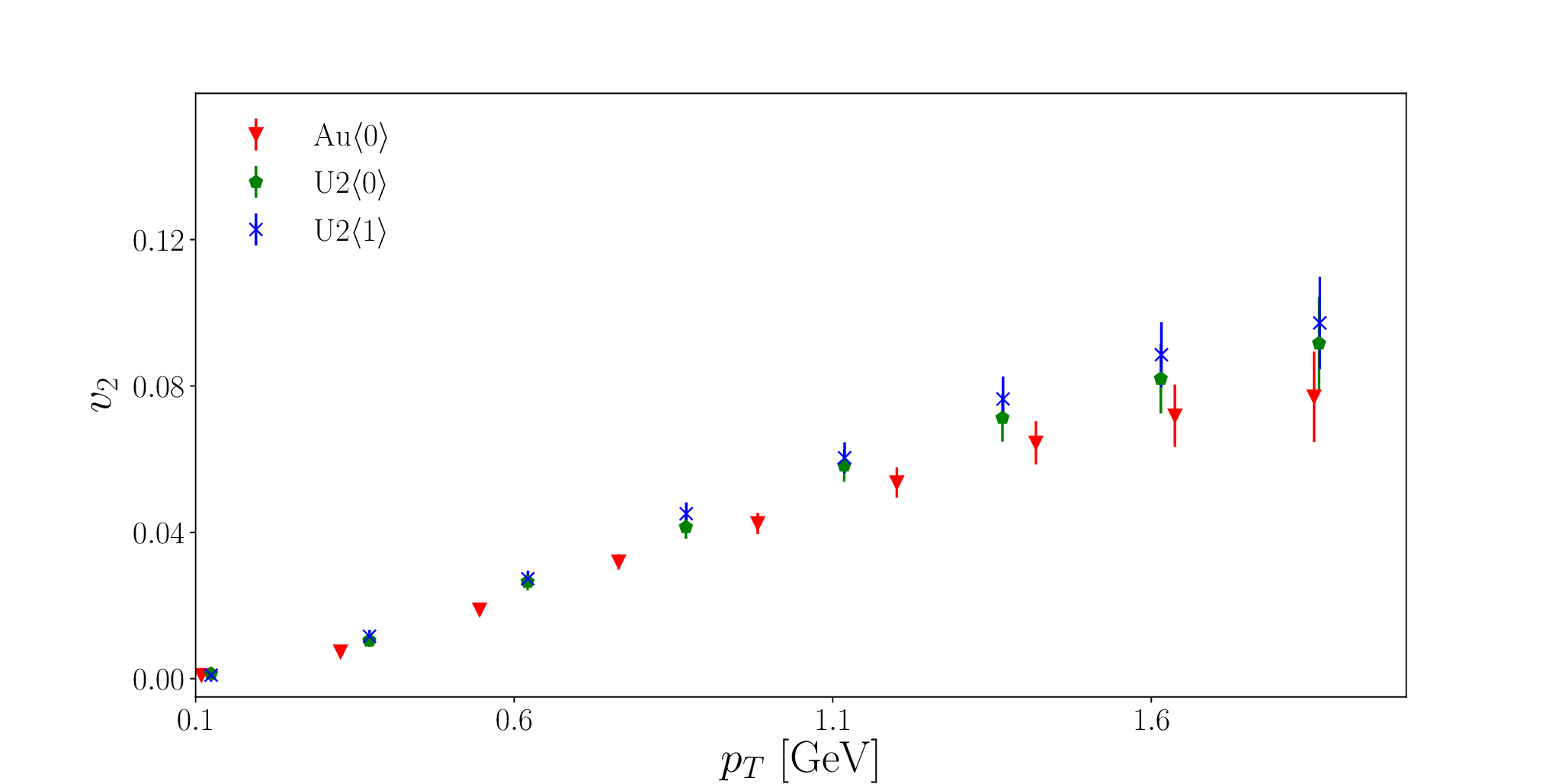}
\includegraphics[width=0.48\linewidth,height=0.3\linewidth]{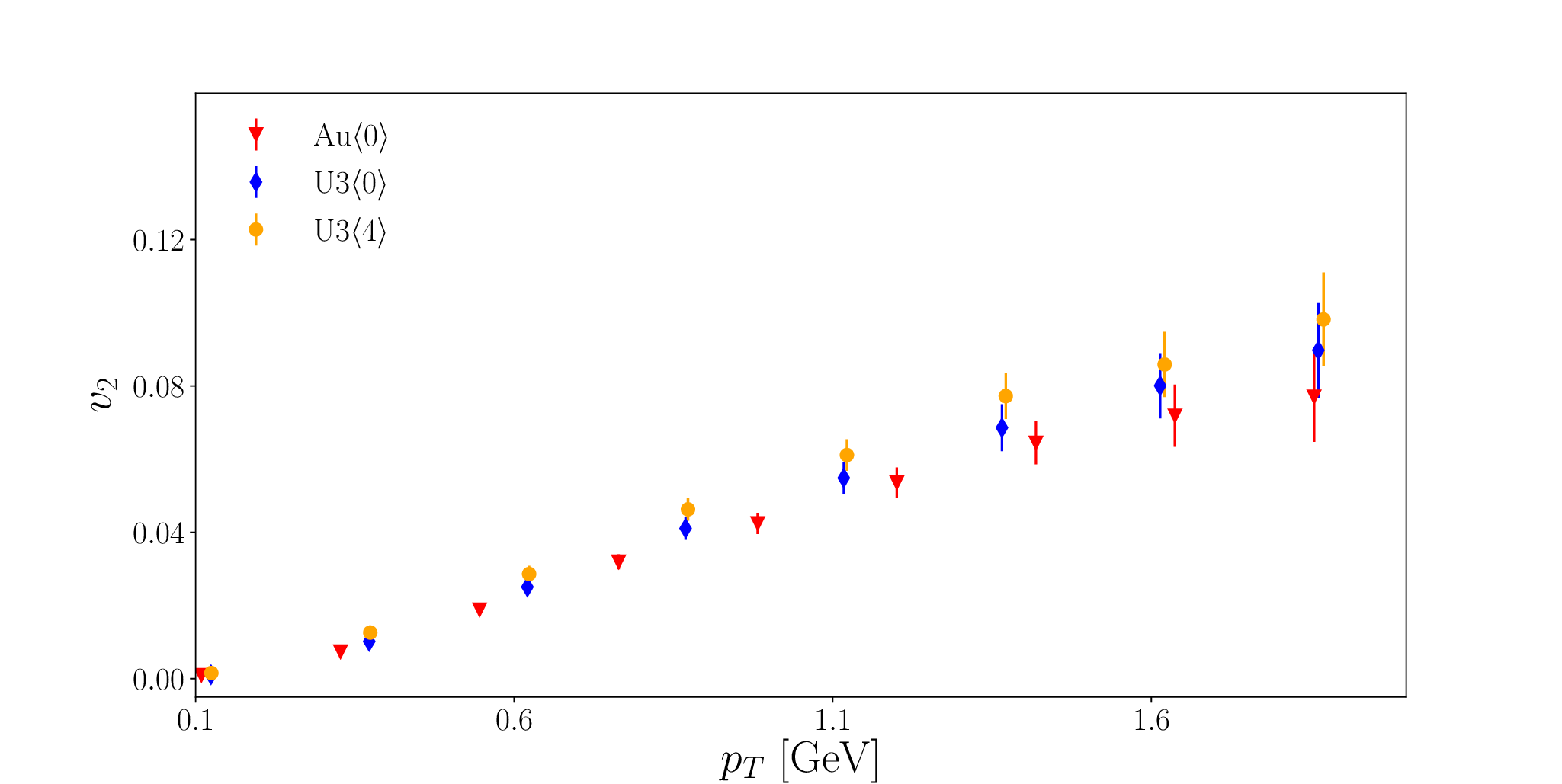}
\includegraphics[width=0.48\linewidth,height=0.3\linewidth]{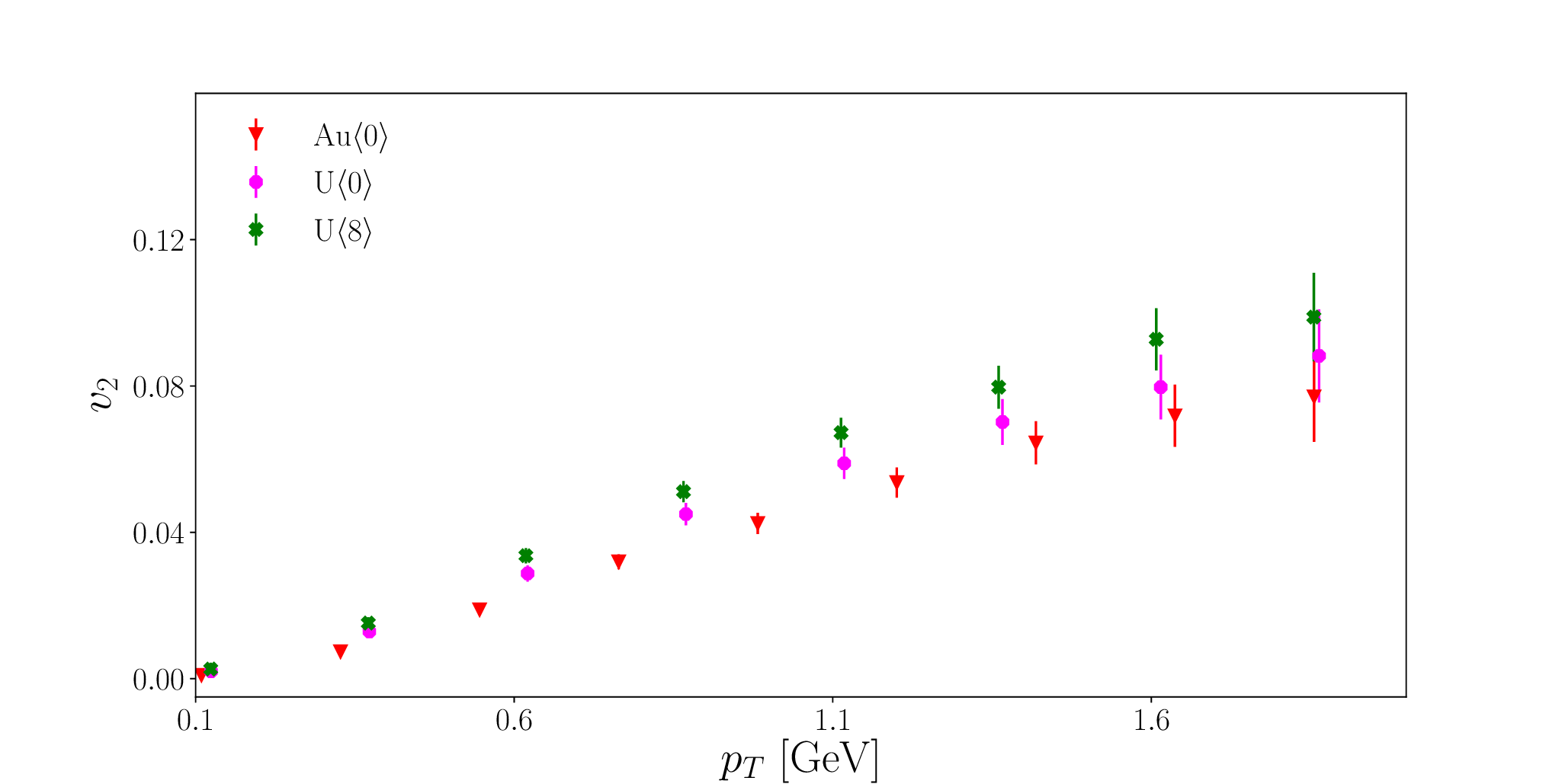}
\includegraphics[width=0.48\linewidth,height=0.3\linewidth]{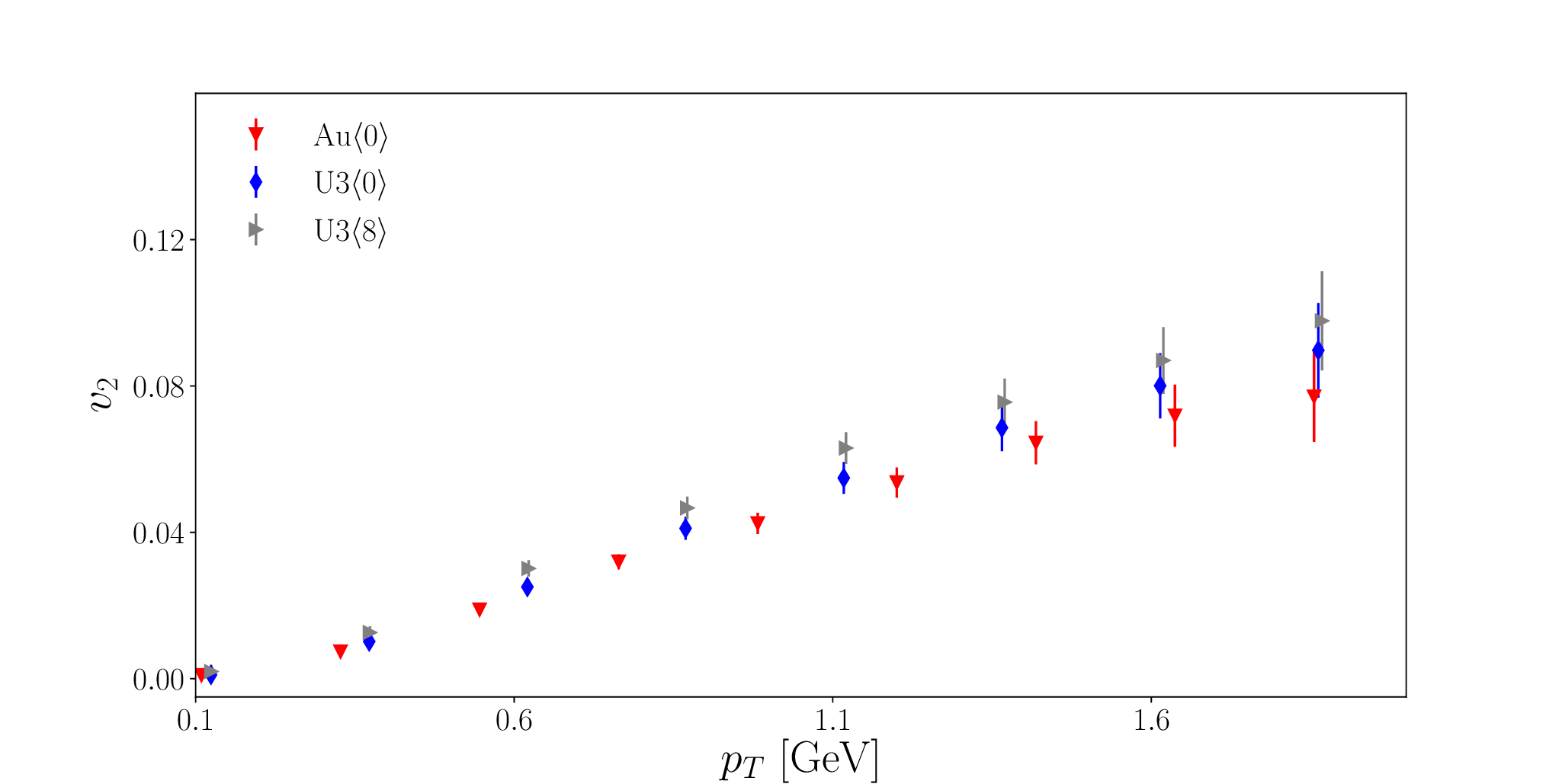}
\caption{Elliptic flow $v_2$ comparison for Au+Au collisions at $\sqrt{s_{NN}}=200$ GeV and different U+U collisions at $\sqrt{s_{NN}}=193$ GeV.
Top-Left: Symmetric Au nuclei compared to symmetric and deformed U nuclei in terms of $\beta_2$.
Top-Right: Symmetric Au nuclei compared to symmetric and deformed U2 nuclei.
Middle-Left: Symmetric Au nuclei compared to symmetric and deformed U3 nuclei in terms of $\beta_2$.
For all the plots, the red triangles are used to represent the results for symmetric Au nuclei.
Middle-Right: Symmetric Au nuclei compared to symmetric and deformed U nuclei in terms of $\beta_4$.
Bottom: Symmetric Au nuclei compared to symmetric and deformed U3 nuclei in terms of $\beta_4$.
The error bars indicate the standard errors.}
\label{Au-U-comp}
\end{figure}

As one can see, comparing Au+Au to U+U data, the elliptic flow for Au+Au collisions is underneath U+U data for the three types of U nuclei.
In the top-left of Fig.~\ref{Au-U-comp}, the elliptic flow of Au+Au collisions is compared to the elliptic flow of U+U collisions.
The top-right figure of Fig.~\ref{Au-U-comp} indicates the results of comparing the elliptic flow from Au+Au to U2+U2 collisions.
At the middle-left of Fig.~\ref{Au-U-comp} we compare the $v_2$ of Au+Au and U3+U3 collisions.
These figures are intrinsically related to the deformation factor $\beta_2$ effect. 
We also compare the results obtained for deformed uranium nuclei according to variations in the $\beta_4$ parameter.
In the middle-right of Fig.~\ref{Au-U-comp} we compared the symmetric Au+Au collisions to symmetric and deformed U+U collisions in terms of $\beta_4$ parameter.
Finally, at the bottom of Fig.~\ref{Au-U-comp} the symmetric Au+Au collisions data is compared to U3+U3 collisions in terms of $\beta_4$ variations.

We can observe similar behavior for elliptic flow for $p_T<0.6$ GeV.
In this regime, Au, U, U2, and U3 essentially assume the same form, while for $p_T>0.6$ GeV, the results computed for Au+Au, U+U, U2+U2, and U3+U3 differ, with uranium collisions presenting high $v_2$.
These results were found for different degrees of deformation in terms of parameters $\beta_2, \,\beta_4$.
The same behavior presented in our study was found at~\cite{Giacalone}.

Last but not least, we discuss the effect of deformation on elliptic flow in the low transverse momentum region.
To this end, we evaluate the average $v_2$ as a function of deformation parameters $\beta_2$ and $\beta_4$.
As the average elliptic flow receives contribution mainly from the lower $p_T$ region, the calculations are primarily aimed at the relevant region where the dependence on transverse momentum is more sensitive. 
When one of the deformation parameters varies for a given nucleus, the remaining parameters are held constants by their original values given in Tab.~\ref{trento_imple-defor}.
The results are presented in Fig.~\ref{au-av2-b2}, where both Au+Au and U+U collisions have been considered.

\begin{figure}[H]
\centering
\includegraphics[width=0.48\linewidth,height=0.3\linewidth]{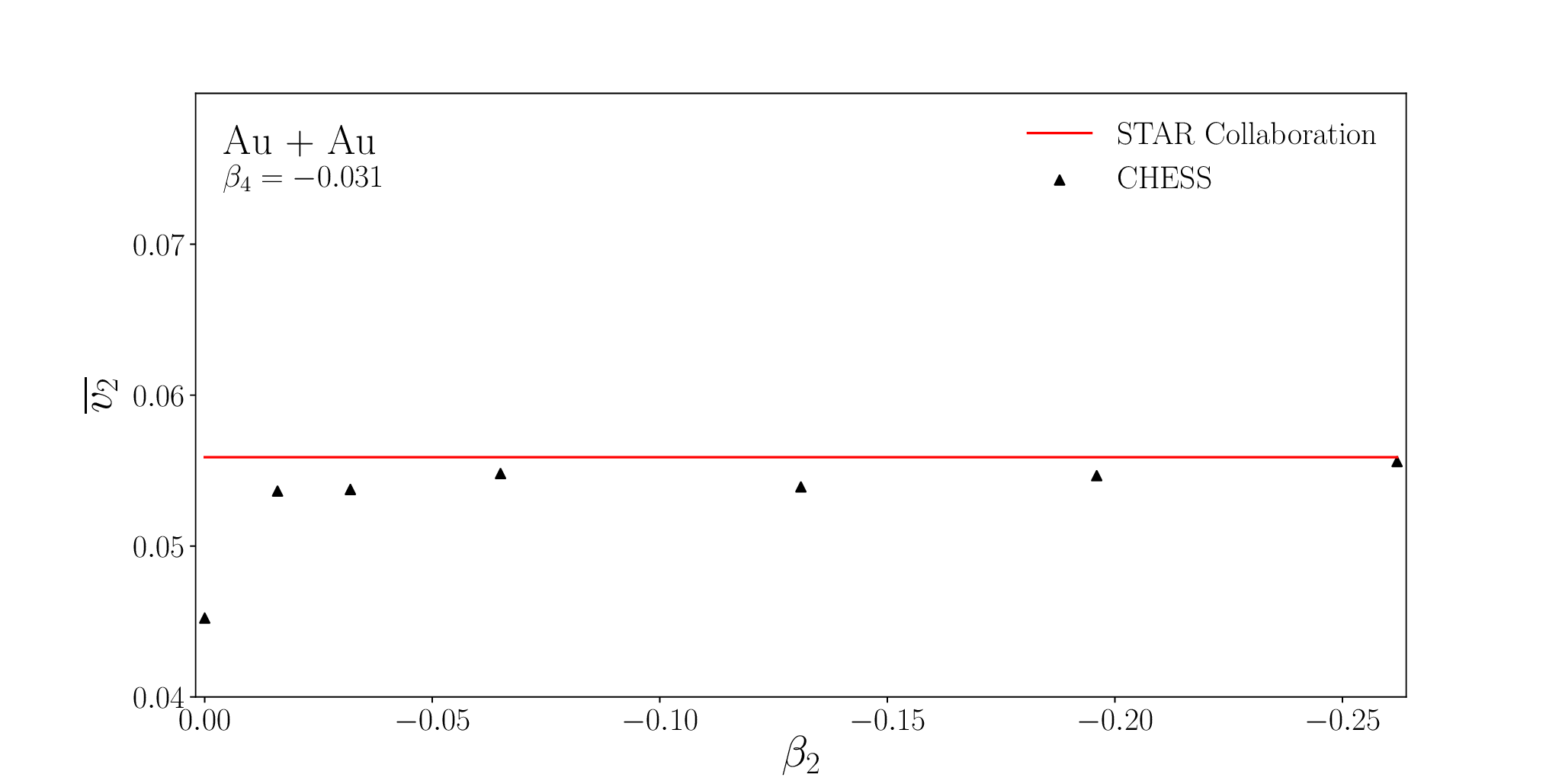}
\includegraphics[width=0.48\linewidth,height=0.3\linewidth]{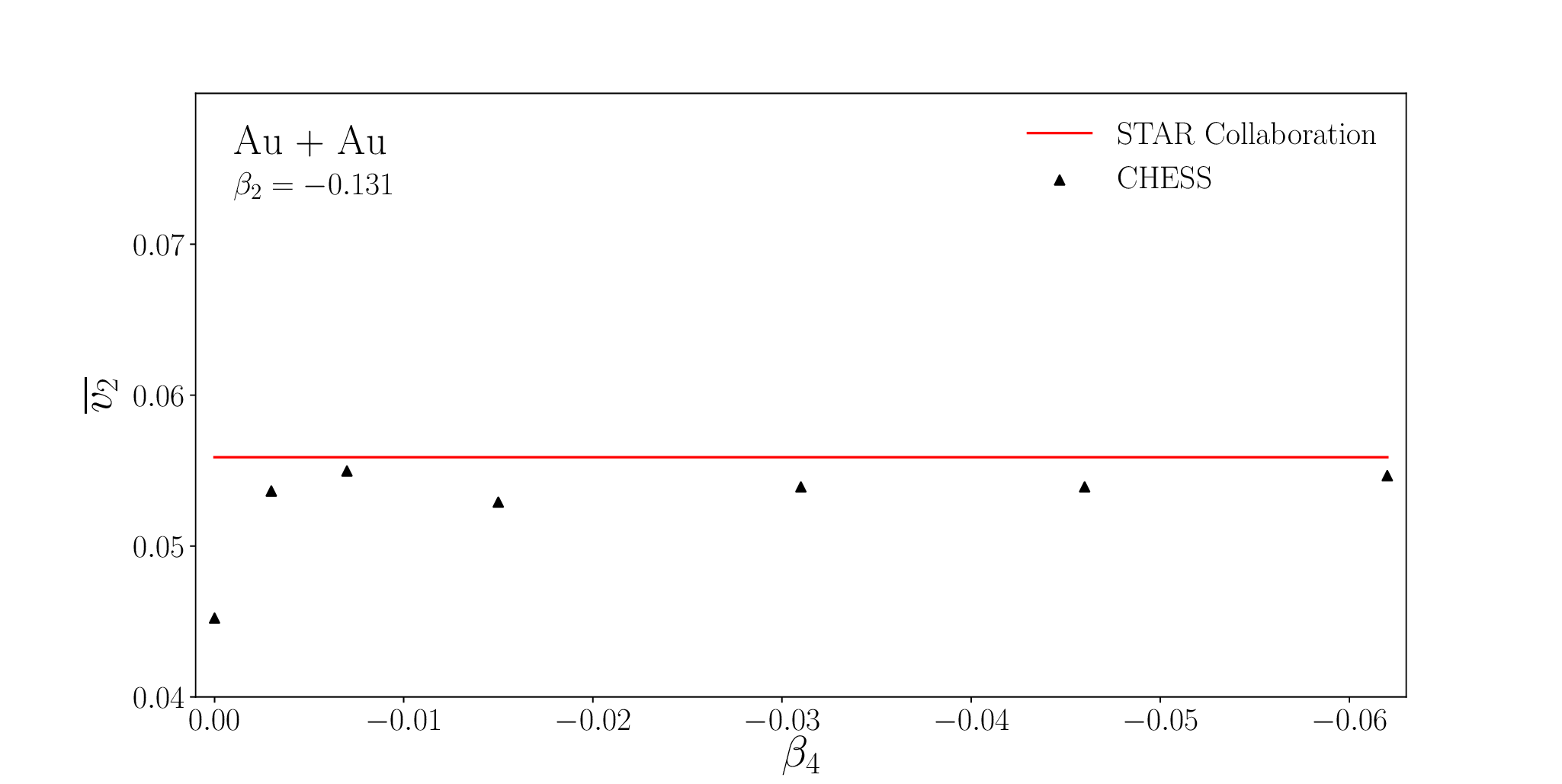}
\includegraphics[width=0.48\linewidth,height=0.3\linewidth]{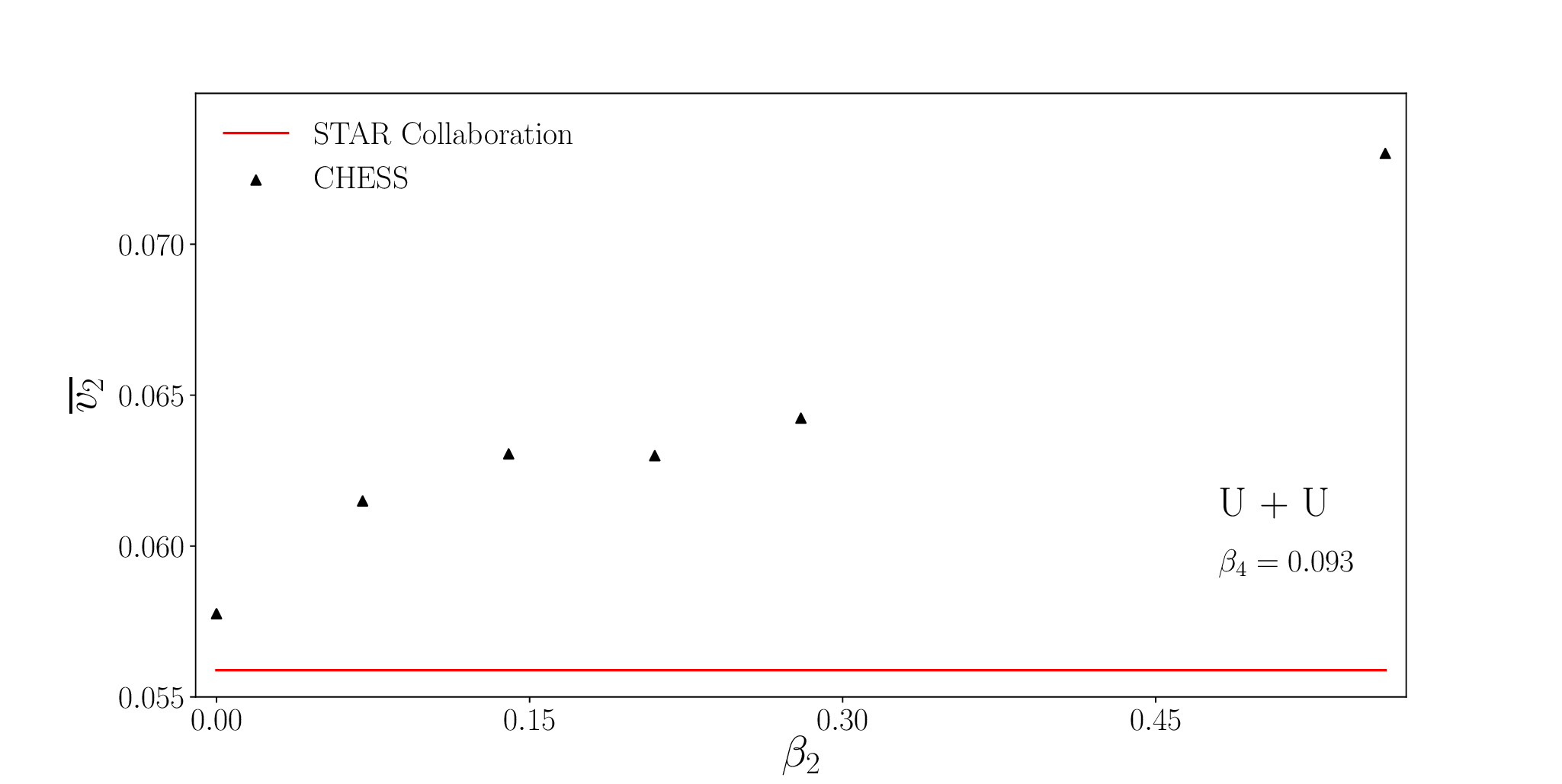}
\includegraphics[width=0.48\linewidth,height=0.3\linewidth]{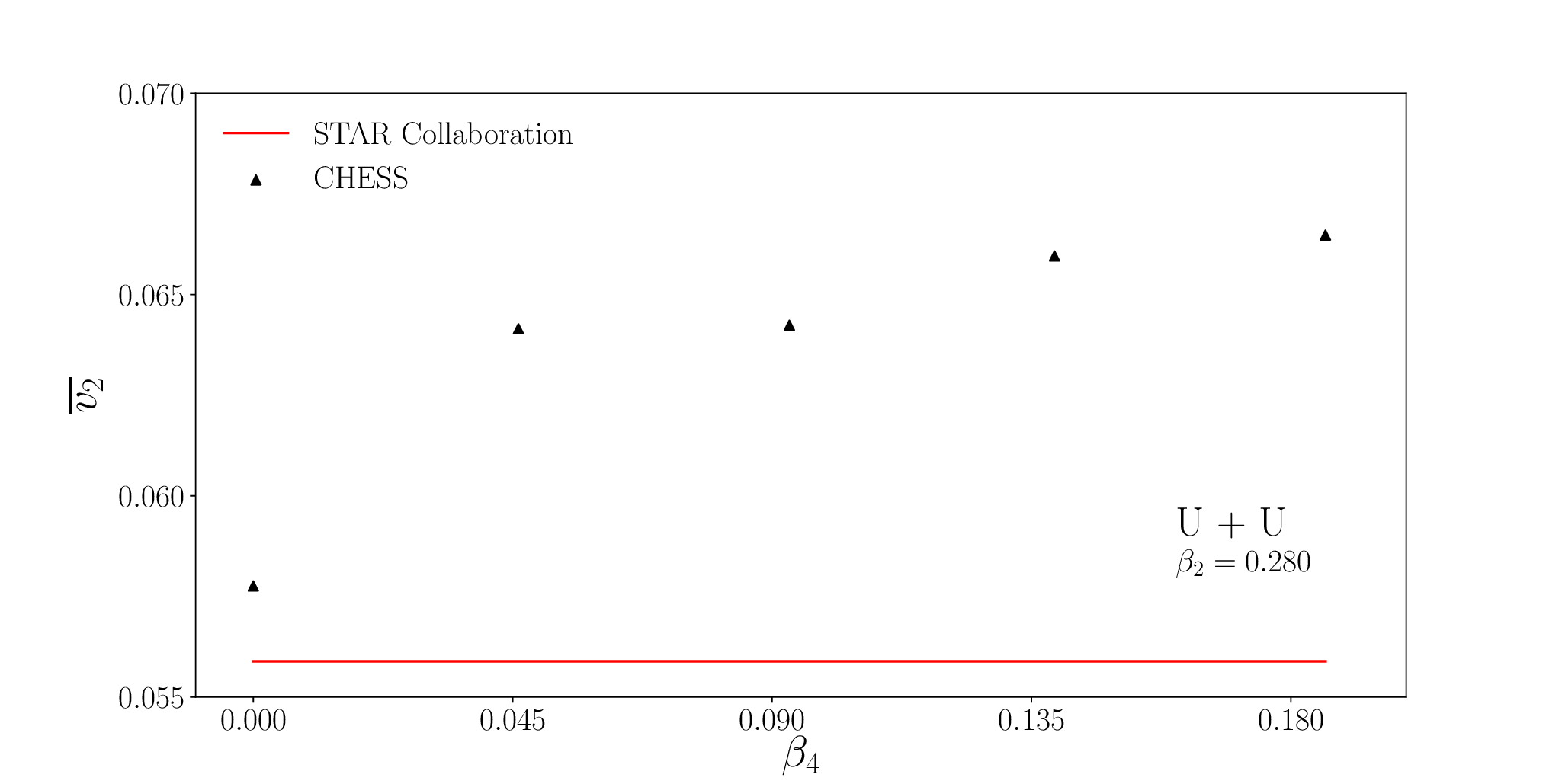}
\includegraphics[width=0.48\linewidth,height=0.3\linewidth]{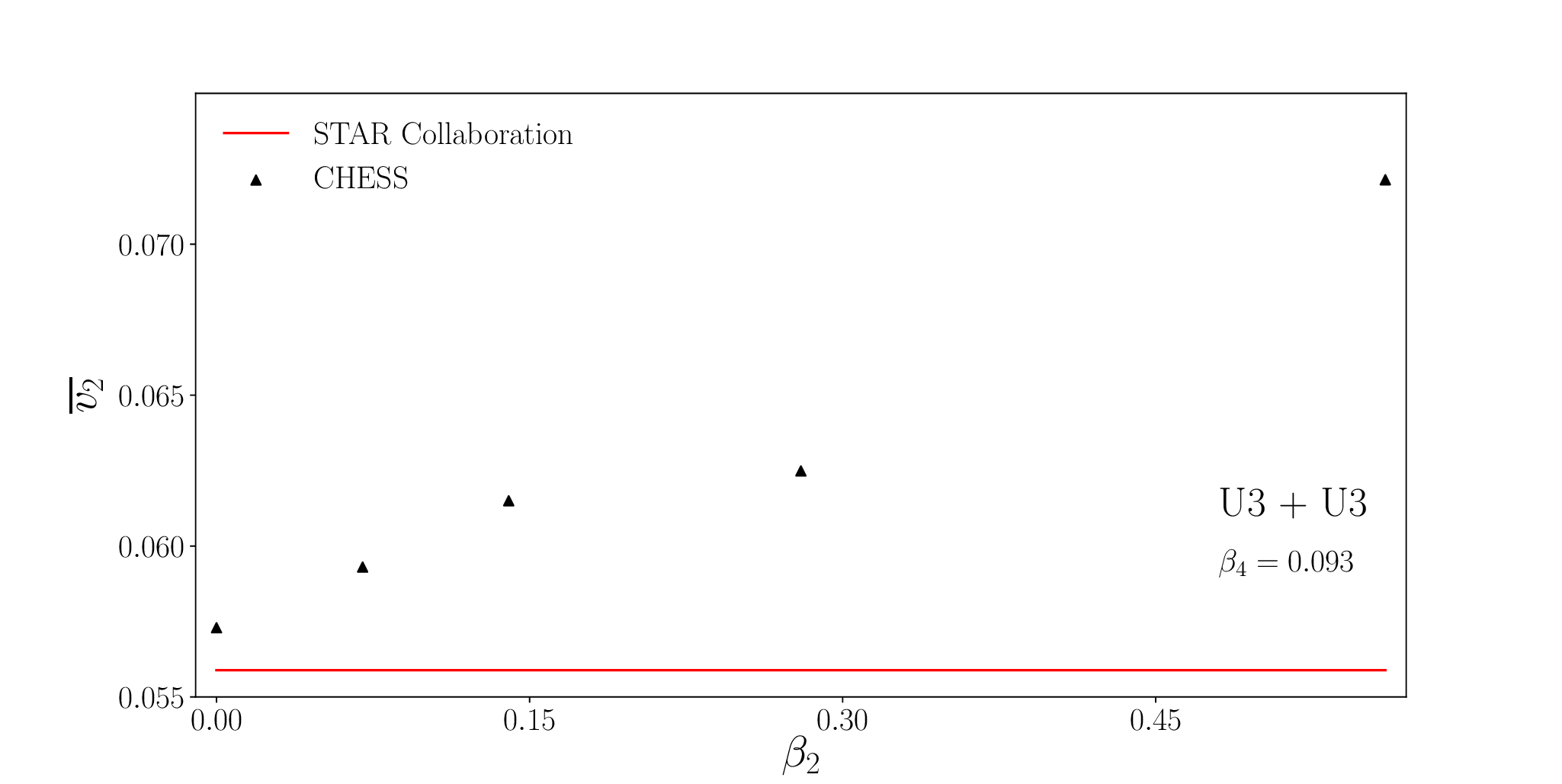}
\includegraphics[width=0.48\linewidth,height=0.3\linewidth]{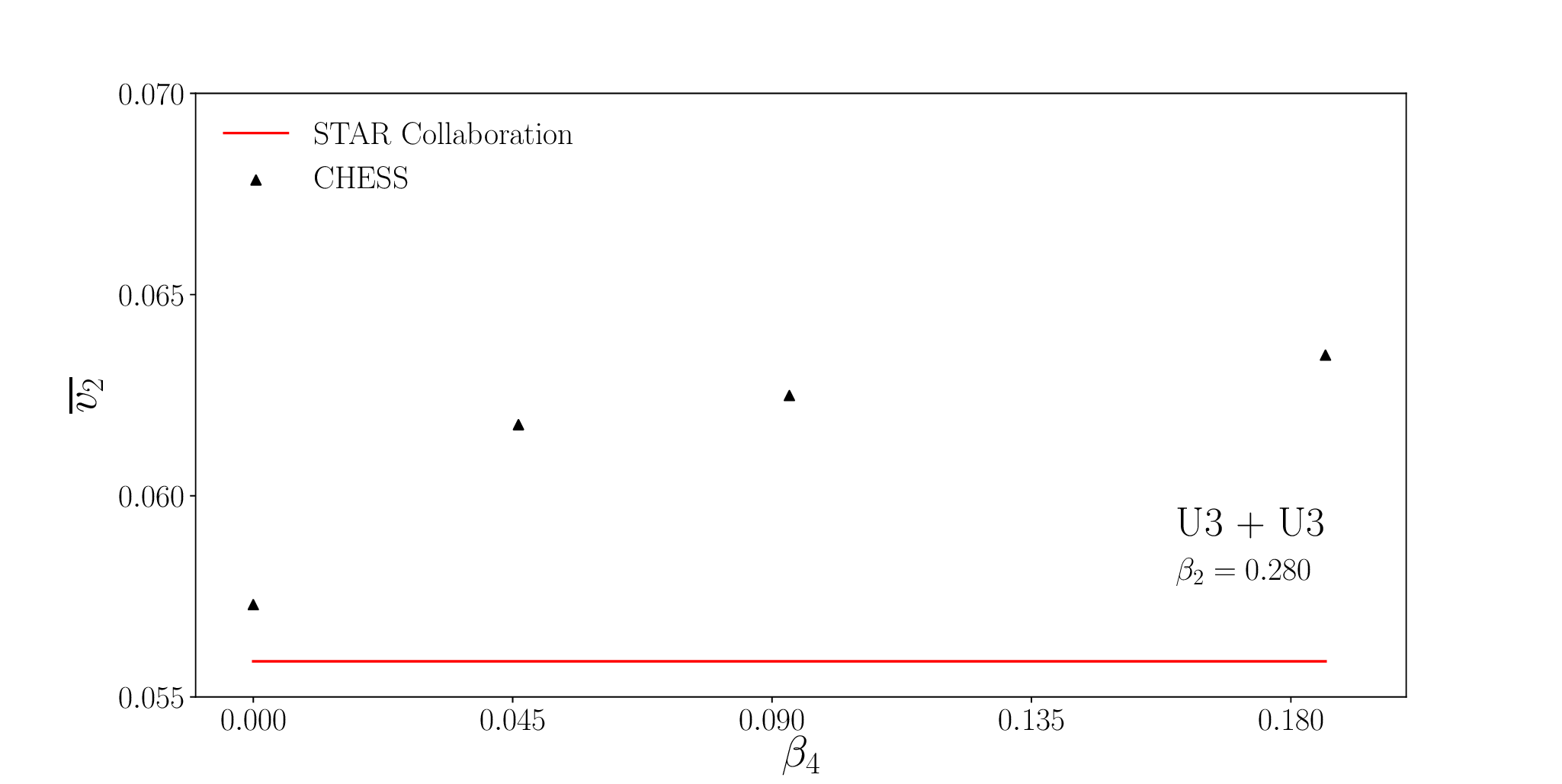}
\caption{Average elliptic flow $v_2$ for Au+Au and U+U collisions with centrality 0-6\% as a function of nucleus deformation parameter.
The numerical results obtained by the CHESS code are compared with the STAR data~\cite{Adams2005}.
Top-Left: The resulting flow as a function of $\beta_2$ for Au+Au collisions.
Top-Right: The resulting flow as a function of $\beta_4$ for Au+Au collisions.
Middle-Left: The resulting flow as a function of $\beta_2$ for U+U collisions.
Middle-Right: The resulting flow as a function of $\beta_4$ for U+U collisions.
Bottom-left: The resulting flow as a function of $\beta_2$ for U3+U3 collisions.
Bottom-right: The resulting flow as a function of $\beta_4$ for U3+U3 collisions.
The specific parameters are given in Tab.~\ref{trento_imple-defor} for all the plots.
The red horizontal line indicates the average flows obtained by STAR Collaboration.}
\label{au-av2-b2}
\end{figure}

For Au+Au collisions, it is observed that the elliptic flow is rather sensitive when the deformation is small. 
When compared with Fig.~\ref{def-perc}, it seems that the somewhat outstanding value of $\bar{v_2}$ is primarily due to the contribution from the anisotropy of low $p_T$ particles.
Here, the average is taken among all the emitted particles and on an event-by-event basis.
As the deformation increases, the impact on the elliptic flow primarily follows a linear relation.
The elliptic flow does not go to zero at vanishing $\beta_2$, which is largely attributed to non-vanishing geometric fluctuations.
For the present study, we found that the value $\beta_2\approx-0.26$ gives the average elliptic flow that is in reasonable agreement with the experimental data.
Additionally, regarding Fig.~\ref{au-av2-b2}, the value $\beta_4 \approx-0.031$ provides the best fit to the experimental data, which was originally adopted by $T_RENT_o$.
As a comparison, Giacalone~\cite{GiacaloneB} found a relatively smaller value $\beta_2\approx-0.15$, by considering only the quadrupole deformation.
We attribute such a difference mostly to that of the hydrodynamic model and IC.

For U+U collisions, the average values of $\bar{v_2}$ show very similar behavior to those for the differential flow, analyzed above in Fig.~\ref{U-beta-full-b2}.
This is expected, as the differential flows are found to be monotonical functions of transverse momentum, and the magnitudes of the effect are also monotonically increasing with increasing deformation parameters.
When comparing the effect of $\beta_2$ and that of $\beta_4$, the latter is found to play a slightly more significant role.

\section{Concluding remarks}\label{section4}

In this work, we explore the sensitivity of the nucleus deformation on flow harmonics using a hybrid code CHESS.
By implementing the initial quadrupole and hexadecapole geometric deformations into the Woods-Saxon profile through $T_RENT_o$, viscous hydrodynamic evolutions are simulated for Au+Au and U+U collisions.
The particle spectrum and collective flow are evaluated and explored.
In particular, we analyze the collective flow as a function of the deformation parameters $\beta_2$ and $\beta_4$. 

Our numerical results prompt a comparison between Au+Au and U+U nuclear collisions, revealing subtle differences in the resulting behavior. 
In the case of Au+Au collisions, the simulations for both symmetric and deformed nuclei give reasonable results regarding the experimental data from the STAR Collaboration, as shown in Fig.~\ref{def-perc}. 
The results for deformed nuclei with slightly augmented flow harmonics stay closer to the experimental data. 
On the other hand, as shown in Fig.~\ref{au-av2-b2}, the transition from symmetric to deformed nuclei marks a shift in average elliptic flow, $\overline{v_2}$, owing to the contributions from high transverse momentum $p_T\sim 2$ GeV. 
For symmetric nuclei, $\overline{v_2}$ significantly overestimates the data, and as deformation increases, the average elliptic flow converges to the experimental data.
Conversely, for U+U nucleus collisions, numerical simulations diverge further from the STAR data as deformation factors increase. 
These results are explicitly illustrated in Fig.~\ref{U-beta-full-b2}, where the elliptic flow increases in response to the deformation factors. 
The computation of the average elliptic flow, presented in Fig.~\ref{au-av2-b2}, supports the linear behavior described in Eq.~\eqref{ani-final-flow} that is extensively documented in the literature.

In the literature, the studies regarding the effect of initial geometric deformation are mainly motivated to answer the following question: whether such deformation is observable and which deformation factor, $\beta_2$ or $\beta_4$, predominantly influences collective flow? 
From the present study, the answer remains elusive, aligned with the results obtained by other authors~\cite{Zhang-Jia, Giacalone, GiacaloneA, GiacaloneB}.
Nonetheless, for collisions of almost symmetric Au nuclei, the resulting integrated flow is found to be somewhat sensitive to the initial quadrupole and hexadecapole deformations.
In this regard, it is interesting to scrutinize the topic by employing advanced analysis tools, such as Bayesian analysis~\cite{hydro-ml-bayesian-05, hydro-ml-bayesian-06, hydro-ml-bayesian-23} and maximum likelihood estimator~\cite{sph-vn-10, sph-vn-11}, and in particular emulator~\cite{hydro-ml-bayesian-22, hydro-ml-bayesian-29, hydro-ml-bayesian-30} and variational autoencoder, which have demonstrated their significant potential recently~\cite{agr-ml-bayesian-05,agr-ml-bayesian-17,agr-ml-bayesian-20,agr-ml-bayesian-22}.

\bibliographystyle{h-physrev}
\bibliography{references_mascalhusk, references_qian}

\end{document}